\documentclass[12pt]{article}
\usepackage[dvips]{graphicx}
\usepackage{epsfig}
\usepackage{latexsym}
\usepackage{amssymb}
\usepackage{amsmath}

\usepackage{wrapfig}
\usepackage{boxedminipage}
\usepackage{setspace}
\usepackage{subfigure}
\DeclareSymbolFont{AMSb}{U}{msb}{m}{n}
\DeclareMathSymbol{\IN}{\mathbin}{AMSb}{"4E}
\DeclareMathSymbol{\IZ}{\mathbin}{AMSb}{"5A}
\DeclareMathSymbol{\IR}{\mathbin}{AMSb}{"52}
\DeclareMathSymbol{\Q}{\mathbin}{AMSb}{"51}
\DeclareMathSymbol{\II}{\mathbin}{AMSb}{"49}
\DeclareMathSymbol{\IC}{\mathbin}{AMSb}{"43}
\DeclareMathSymbol{\IP}{\mathbin}{AMSb}{"50}
\DeclareMathSymbol{\IH}{\mathbin}{AMSb}{"48}
\DeclareMathSymbol\IA{\mathalpha}{AMSb}{"41}
\DeclareMathSymbol\IS{\mathalpha}{AMSb}{"53}

\def\Q{{\cal Q}}

\begin{document}

\bigskip
\bigskip
\bigskip
\bigskip
\bigskip

\begin{center} {\Large \bf  Meson Spectra and Magnetic Fields}
  
\bigskip

{\Large\bf  in the }

\bigskip

{\Large\bf  Sakai--Sugimoto Model}

\end{center}

\bigskip \bigskip \bigskip \bigskip

\centerline{\bf  Clifford V. Johnson, Arnab Kundu}

\bigskip
\bigskip
\bigskip

\centerline{\it Department of Physics and Astronomy }
\centerline{\it University of
Southern California}
\centerline{\it Los Angeles, CA 90089-0484, U.S.A.}

\bigskip

\centerline{\small \tt johnson1, akundu  at usc.edu}

\bigskip
\bigskip

\begin{abstract} 
\noindent 
We continue our study of the dynamics of the flavour sector of the
Sakai--Sugimoto model in the presence of an external magnetic field,
uncovering several features of the meson spectrum at high and low
temperatures. We employ both analytical and numerical methods to study
the coupled non--linear equations that result from the gravity dual.
\end{abstract}

\newpage \baselineskip=18pt \setcounter{footnote}{0}

\section{Introduction}

Gauge/gravity dualities have become powerful tools for understanding
diverse phenomena in strongly coupled gauge theories. In this paper we
continue a line of investigation concerning the influence of an
external magnetic field on the dynamics of the flavour sector of a
gauge theory. The ultimate goal is to try to isolate features that are
of interest for learning about the properties of nuclear matter, the
idea being that these techniques provide an alternative (and/or
complementary) route to that of Quantum Chromodynamics (QCD) for
access to the physics of interest in various regimes. The model we
study here has no pretensions to being close in microscopic detail to
the known strong interaction physics (such as those captured by QCD),
but nevertheless there is good reason to believe that there are
phenomena that can be captured using it that are not sensitive to all of the
detail.

The model that we study here was, in essence, constructed by
Witten in ref.~\cite{Witten:1998zw} by taking the near--horizon limit
of D4--branes compactified on a spatial circle to capture the physics
of confinement/deconfinement transition in QCD. This model was later
revisited in refs.~\cite{Sakai:2004cn, Sakai:2005yt} incorporating the
dynamics of probe D8/anti--D8 flavour branes, which led to a geometric
understanding of spontaneous chiral symmetry breaking. It is now
commonly known as the Sakai--Sugimoto model.

In the Sakai--Sugimoto model the dynamics of the flavour sector has
been analyzed at finite temperature in~refs.~\cite{Aharony:2006da,
  Parnachev:2006dn} and in the presence of an external electromagnetic
field in refs.~\cite{Johnson:2008vna, Bergman:2008sg}. For magnetic
fields, it has previously been argued in {\it e.g.}
ref.~\cite{Miransky:2002eb} that for $(3+1)$ and $(2+1)$-dimensional
field theories there is a universal phenomenon at work that is
sometimes referred to as magnetic catalysis of chiral symmetry
breaking.  These field theory treatments typically involve analyzing
truncated Dyson--Schwinger equations in order to get access to the
non--perturbative physics.

This physics (that of a magnetic field catalyzing chiral symmetry
breaking) can now be studied using the alternative technique of
gauge/gravity duals (the magnetic field maps to a background B--field
in the string theory and causes the flavour branes in the geometry to
bend, generically), and it has been observed\cite{Johnson:2008vna} to
persist in all the Sakai--Sugimoto type models, where the flavour
degrees of freedom are inserted as defects in the dual geometry for
several $(2+1)$-- and $(3+1)$--dimensional gauge theories. A similar
effect has been observed to exist in other models in
refs.~\cite{Filev:2007gb, Albash:2007bk}.  In those works, some of the
structure of the meson spectrum was also studied, revealing phenomena
such as Zeeman splitting,  level mixing, {\it etc}. Here we would like
to study the meson spectrum in the presence of an external magnetic field
in the Sakai--Sugimoto model supplementing the study of the phase
diagram obtained in our earlier work\cite{Johnson:2008vna}.

Previous studies of the meson spectrum in this particular model have
been carried out in {\it e.g.}, ref. \cite{Peeters:2006iu} for both
low spin and high spin mesons. We uncover the role of the magnetic
field in these spectra. For the low spin mesons we restrict ourselves
to the high temperature phase only and therefore study the quasinormal
modes of the scalar and a subset of the vector mesons. This comes with
a caveat. The precise identification between the quasinormal frequency
associated with the supergravity fluctuations in this model and the
meson spectrum in the dual field theory is rather subtle and poorly
understood\footnote{For a detailed analysis regarding the issue of
  precise identification, see ref.~\cite{Paredes:2008nf}.}. In the
Sakai--Sugimoto type models the mesons transform under an $U(N_f)_{\rm
  diag}$, whereas the quasinormal modes transform under either an
$U(N_f)_L$ or an $U(N_f)_R$. This fact alone readily makes it
difficult to precisely define what gauge theory quantities we would be
studying by extracting the quasinormal modes. Nevertheless the
quasinormal modes are naturally associated to the embeddings falling
into the black hole of the gravity dual and therefore we study those
in their own rights.

Moreover, since the embedding function of the probe D8--brane is
trivial ($\tau_0(u)={\rm const}$, where $\tau_0(u)$ denotes the
profile function) in the high temperature phase, the scalar and the
vector fluctuations remain decoupled even in the presence of a
background anti--symmetric field such as a constant magnetic
field. This is to be contrasted with the D3--D7 model in a background
magnetic field\cite{Albash:2007bq} where the profile function of the
probe D7--brane is non-trivial and therefore the scalar fluctuations
couple with the vector fluctuations. Thus studying quasinormal modes
in presence of a background anti--symmetric field in general is more
difficult in the later model.

The study of high spin mesons is another avenue that we pursue. We
find that the presence of the magnetic field enhances the stability of
such mesons by increasing their angular momentum. This is turn increases
the dissociation temperature at which they fall apart into their
constituents. We also uncover a simple realization of the Zeeman
effect in the presence of a background field. It can be readily seen
that such effects exist in all the Sakai--Sugimoto type models with
probe branes of diverse dimensions.

This article is divided into different sections. In section 2, we
review the Sakai--Sugimoto model and its phase diagram in presence of a
magnetic field. We then move on to section 3 to study the quasinormal
modes for scalars and the decoupled sector of transverse and
longitudinal vector mesons in this model. Section 4 is devoted to the
study of high spin mesons, where we first  obtain some
analytical results and then proceed to the numerical analysis. We have
relegated the details of some of the calculations to four
appendices. Appendix A contains the details of the Lagrangian for the
quadratic fluctuations and the equations of motion for the gauge
fields as well as a detailed discussion about when they can be
decoupled. In appendix B we study the longitudinal and transverse
modes in the hydrodynamic limit in vanishing magnetic field and obtain
the well--known hydrodynamic dispersion relation for the longitudinal
mode. Appendix C contains the general variable changes we use to
recast the equations of motion for the fluctuation modes into an
effective Schr\"{o}dinger equation. In appendix D we present a
model calculation of high spin mesons in a background magnetic field
by analyzing spinning strings in Rindler space coupled to a magnetic
field. This serves as a toy model that captures some of the features
of meson dissociation in a background field.

\section{Review of the Model}

The Sakai--Sugimoto model is constructed from the near horizon limit of $N_c$ $D4$-branes with one compact spatial direction along which an anti-periodic boundary condition is imposed for the adjoint fermions. This makes the adjoint fermions massive and breaks supersymmetry. Flavours are introduced by $N_f$ $D8$ and $\overline{D8}$-branes intersecting the $D4$-brane at a $(3+1)$-dimensional defect. Thus there is a global $U(N_f)_L\times U(N_f)_R$ flavour symmetry as viewed from the worldvolume of the $D4$-brane. We work in the probe limit, namely $N_f\ll N_c$ where the flavour branes do not backreact on the background.

At low temperature the near horizon geometry of the $D4$-brane is given by 
\begin{eqnarray}
&& ds^2 = \left(\frac{u}{R}\right)^{3/2}\left(dt_E^2+dx_idx^i+f(u)d\tau^2\right)+\left(\frac{u}{R}\right)^{-3/2}\left(\frac{du^2}{f(u)}+u^2d\Omega_4^2\right)\ ,   \nonumber\\
&&   e^{\phi}=g_s\left(\frac{u}{R}\right)^{3/4}\ , \quad F_{(4)}=\frac{2\pi N_c}{V_4}\epsilon_4\ ,\quad f(u)=1-\left(\frac{u_{KK}}{u}\right)^3\ , \quad R^3=\pi g_s N_c \ell_s^3\ ,
\end{eqnarray} 
where $x^i$ are the flat $3$-directions, $t_E$ is the Euclidean time coordinate, $\tau$ is the spatial compact circle, $d\Omega_4^2$ is the metric on the round $S^4$ and $u$ is the radial direction, $\ell_s$ is the string length, $g_s$ is the string coupling; $V_4$ and $\epsilon_4$ are the volume and volume form of $S^4$ respectively. Also, $\phi$ is the dilaton and $F_{(4)}$ is the RR four-form field strength. The radius of the compact spatial circle is given by $2\pi R_4=(4\pi R^3)/(3u_{KK}^{1/2})$, which is obtained from the condition that the compact direction $\tau$ shrinks away smoothly at $u=u_{KK}$. 

The five-dimensional gauge coupling is given by $g_5^2=(2\pi)^2g_s\ell_s$, the four dimensional gauge coupling can be obtained by dimensional reduction yielding $g_4^2=g_5^2/2\pi R$ and the five dimensional 't Hooft coupling is defined to be $\lambda=(g_5^2 N_c)/4\pi$. Gravity calculations are reliable when the spacetime curvature is small compared to the string tension and also when string loop effects are suppressed. Taken together these two conditions tell us that the gravity approach is valid in strong 't Hooft coupling limit of the four-dimensional gauge theory (see {\it e.g.} ref. \cite{Aharony:2006da}).

The high temperature background is given by
\begin{eqnarray}\label{eqt: highmet}
&& ds^2=\left(\frac{u}{R}\right)^{3/2}\left(dx_idx^i+f(u)dt_E^2+ d\tau^2\right)+\left(\frac{u}{R}\right)^{-3/2}\left(\frac{du^2}{f(u)}+u^2d\Omega_4^2\right)\ ,\nonumber\\
&& t=t+\frac{4\pi R^{3/2}}{3u_T^{1/2}}\ ,\quad T=\frac{1}{\beta}=\left(\frac{4\pi R^{3/2}}{3u_T^{1/2}}\right)^{-1}\ ,\quad f(u)=1-\left(\frac{u_T}{u}\right)^3\ ,\\ \nonumber
\end{eqnarray}
where $T$ is the background temperature and all other parameters are as defined before.

We let the probe brane--anti-brane pair stretch along the directions $\{t,x^i,\Omega_4\}$ and have a non trivial profile described by the function $\tau(u)$. In the low temperature phase the background cigar topology in the $\{\tau,u\}$-submanifold forces the brane pair to join at some radial distance $U_0$ breaking the global $U(N_f)_L\times U(N_f)_R$ down to a $U(N_f)_{\rm diag}$. This is then identified with the spontaneous chiral symmetry breaking in the model. The high temperature phase is richer in content. The cigar topology now appears in the $\{t_E,u\}$-submanifold of the background and therefore the brane--anti-brane pair can end on the horizon separately. Depending on the background temperature or the asymptotic separation between the brane--anti-brane pair, there is a chiral symmetry restoring transition.

The background geometry undergoes a confinement/deconfinement transition at a temperature $T_d=1/(2\pi R_4)$, where $R_4$ is the radius of the compact $\tau$-direction. This is the temperature at which the gluons deconfine. When flavours are added, depending on the dimensionless parameter $L/R_4$ there can be a range of temperatures where the gluons are deconfined and chiral symmetry in the flavour sector is spontaneously broken by virtue of the brane--anti-brane pair joining at the radial position $U_0$. If the quark separation $L$ obeys the bound that $L/R_4<0.97$, then the intermediate phase (which is deconfined but chiral symmetry broken) exists. If however $L/R_4>0.97$, then deconfinement and chiral symmetry restoration happens together. Therefore there can exist a separation of scales between the gluon deconfinement and chiral symmetry restoring transitions. This has been discussed in refs.~\cite{Aharony:2006da, Parnachev:2006dn}.

We can also set $R_4\to\infty$ so that the spatial circle direction now becomes a flat extended direction; this particular limit has been studied in detail in refs.~\cite{Antonyan:2006vw, Antonyan:2006qy, Antonyan:2006pg}. It turns out that in this particular limit the dual gauge theory is described by the Nambu-Jona-Lasinio model (when probing with D8-branes) or the Gross-Neveu model (when probing with D6-branes) with a non-local four-fermi interaction. In this case, the asymptotic separation between the brane--anti-brane pair sets the coupling for the four-fermi interaction. Here the so called intermediate temperature phase always exists.

\begin{figure}[!ht]
\begin{center}
\includegraphics[angle=0,
width=0.65\textwidth]{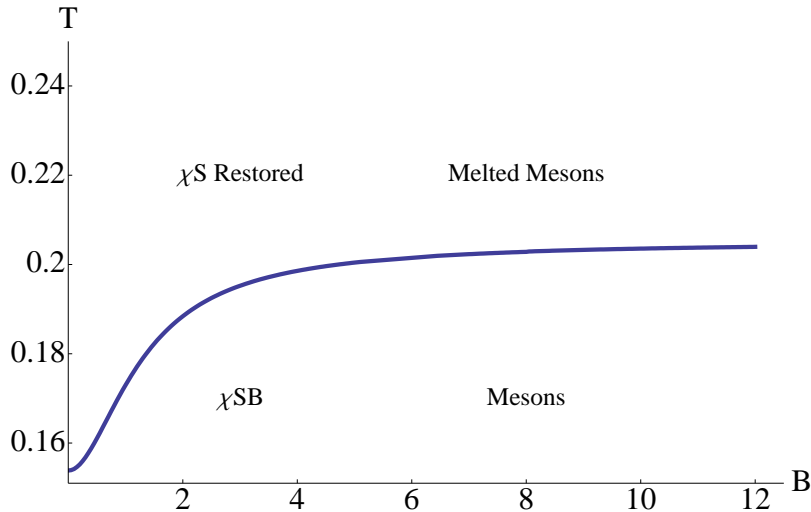}
\caption{\small The phase diagram in $T-B$ plane. The vertical axis is expressed in units of $1/L$ and the horizontal axis is expressed in units of $R$.}
\label{fig: phase}
\end{center}
\end{figure}

In this model, the effect of a background magnetic field on chiral symmetry breaking was analyzed in refs.~\cite{Bergman:2008sg, Johnson:2008vna}. It was explicitly shown that a background magnetic field enhances chiral symmetry breaking and increases the chiral symmetry restoring temperature. The resulting phase diagram is shown in figure \ref{fig: phase}. This magnetic field couples to the flavour sector only, therefore it does not affect the (gluon) confinement/deconfinement transition temperature. Thus the background magnetic field increases the separation of gluon deconfinement and chiral symmetry restoring temperature scales. However, as we can observe from figure \ref{fig: phase} this separation of scales does not increase indefinitely with increasing magnetic field, rather it tends to saturate a maximum value. This particular feature persists in several other Sakai-Sugimoto type models\cite{Johnson:2008vna}.

\section{Mesons with Small Spin}

Mesons with low spin in the dual gauge theory correspond to the small fluctuations of the classical profile of the probe brane. The fluctuation in the geometric shape (which is denoted by the function $\tau(u)$) of the probe branes corresponds to the scalar (or pseudo-scalar) meson mode and the fluctuation of the Maxwell field on the worldvolume of the $D8/\overline{D8}$-brane corresponds to the vector (or axial vector) mesons.

\subsection{Scalar Fluctuation}

Let us consider a fluctuation of the probe brane embedding function $\tau(u)=\tau_0(u)+\chi$, where $\tau_0(u)$ is the classical profile. Before proceeding further let us recall that the function $\tau_0(u)$ is non-trivial only for the U-shaped configuration (for the straight branes, this function is just a constant). Here we will consider the high temperature phase only for which $\tau_0(u)={\rm const.}$ and the scalar fluctuation is always decoupled from the vector fluctuations.

To recast the equation of motion for the fluctuation mode into an equivalent Schr\"{o}dinger equation we define, $f_{\chi}(u)=\sigma(u)g(u)$, where $\sigma(u)$ is defined in eqn. (\ref{eqt: schg}) (the details of the coordinate changes are given in appendix C). The equivalent Schr\"{o}dinger equation is given by (to avoid notational clutter we choose $R=1$ and $U_T=1$)
\begin{eqnarray}\label{eqt: scalar}
&& \partial_{\tilde u}^2g(\tilde u)+\omega^2g(\tilde u)-V_S(\tilde u)g(\tilde u)=0\ , {\rm where} \quad d\tilde{u}=\frac{u^{3/2}}{u^3-1}du\ , \nonumber\\ 
&& {\rm and}\quad V_S(u)=\frac{\left(u^3-1\right) \left(5 B^4 \left(7u^3+5 \right) +2 B^2 u^3 \left(71 u^3+7 \right) +16 u^6
   \left(5 u^3+1\right)\right)}{16 u^5 \left(B^2 +u^3\right)^2}\ .
\end{eqnarray}
Here $\tilde{u}$ is the ``tortoise" coordinate defined in appendix C. For simplicity, we continue to express the Schr\"{o}dinger potential in the original $u$-variable.

We consider only in-falling modes at the horizon; such a solution can be written as $g(\tilde{u})={\rm exp}(-i\omega t)\psi(\tilde{u})$. Now multiplying the first equation in (\ref{eqt: scalar}) by $\psi^*(\tilde{u})$, integrating by parts from the horizon to the boundary and using the equation of motion we get
\begin{eqnarray}\label{eqt: vsch}
\int_{\tilde{u_b}}^\infty d\tilde{u}\left(|\psi'(\tilde{u})|^2+V_s(\tilde{u})|\psi(\tilde{u})|^2\right)=-\frac{|\omega|^2|\psi(\infty)|^2}{{\rm Im}\omega}\ .
\end{eqnarray}

\begin{figure}[!ht]
\begin{center}
\includegraphics[angle=0,
width=0.55\textwidth]{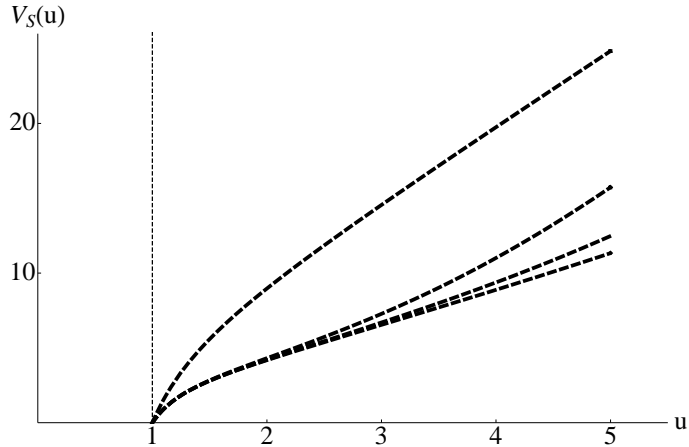}
\caption{\small The Schr\"{o}dinger potential for different values of external magnetic field. The dashed vertical black line represents the position of the horizon, the top-most curve represents the potential at zero external field. As the field increases we find the potential grows more slowly with increasing~$u$.}
\label{fig: vscalarT}
\end{center}
\end{figure}

Now we can understand qualitative features of imaginary part of quasinormal frequency by studying this effective potential. We resort to numerics to study the potential $V_s(u)$. This potential vanishes at $u=1$ because of the presence of the horizon and diverges as $u\to\infty$ reflecting infinite gravitational potential of the background geometry.

The important feature that we find from figure \ref{fig: vscalarT} is that the Schr\"{o}dinger potential is always positive, therefore from eqn. (\ref{eqt: vsch}) we can conclude that the imaginary part of the quasinormal frequency is strictly negative. Moreover, the potential does not develop a negative energy well, therefore there exists no negative energy bound states in this phase. This is consistent with the fact that the probe brane configuration is stable and tachyon free and chiral symmetry restoration is accompanied by meson melting transition. Tachyonic modes in the scalar field fluctuation are likely to appear in the overheated low temperature phase which is bypassed by the transition. Also we see that the potential vanishes at the horizon and therefore the spectrum is continuous, which is consistent with the melting transition.

\begin{figure}[!ht]
\begin{center}
\subfigure[] {\includegraphics[angle=0,
width=0.45\textwidth]{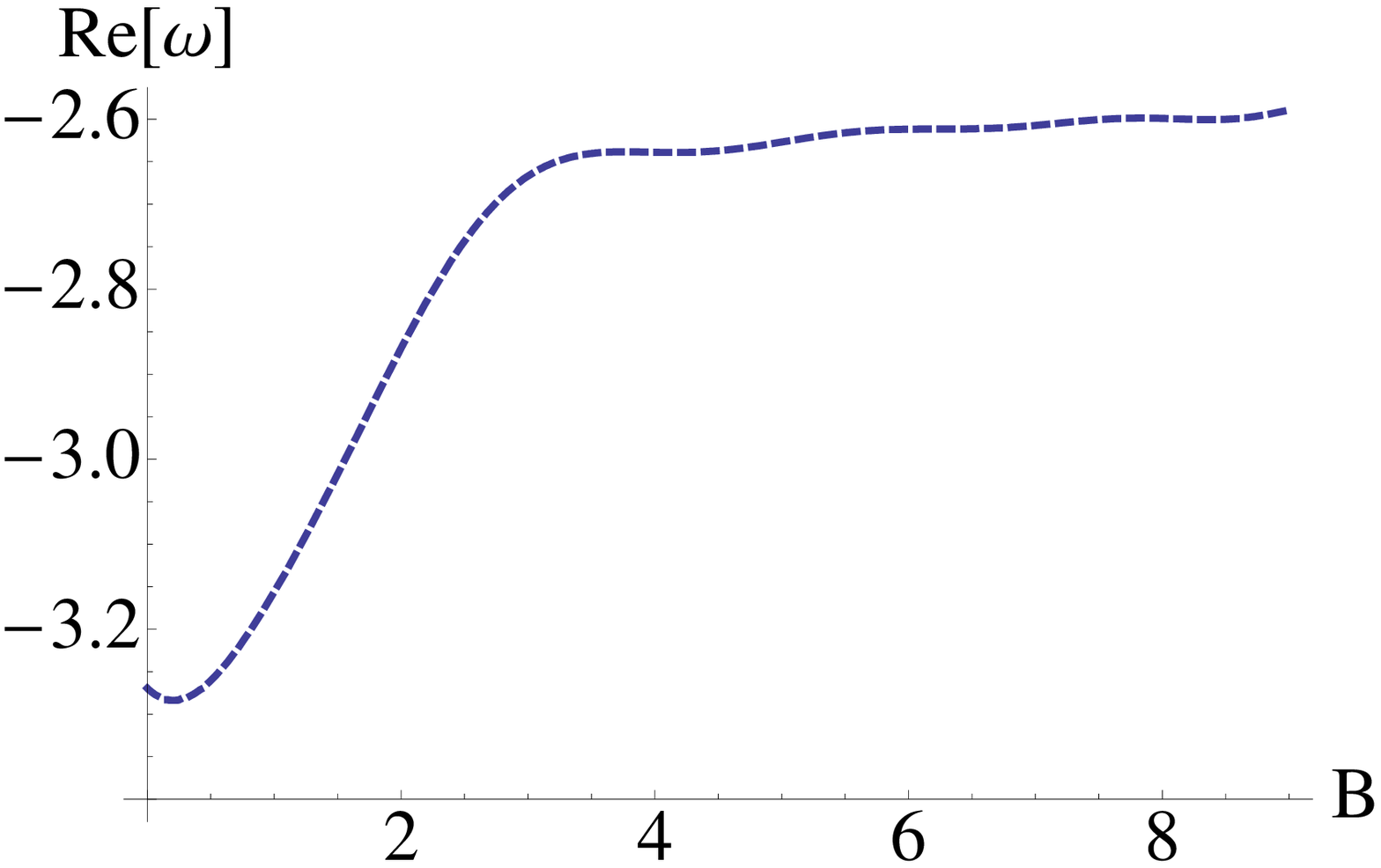} \label{fig: sR}}
\subfigure[] {\includegraphics[angle=0,
width=0.45\textwidth]{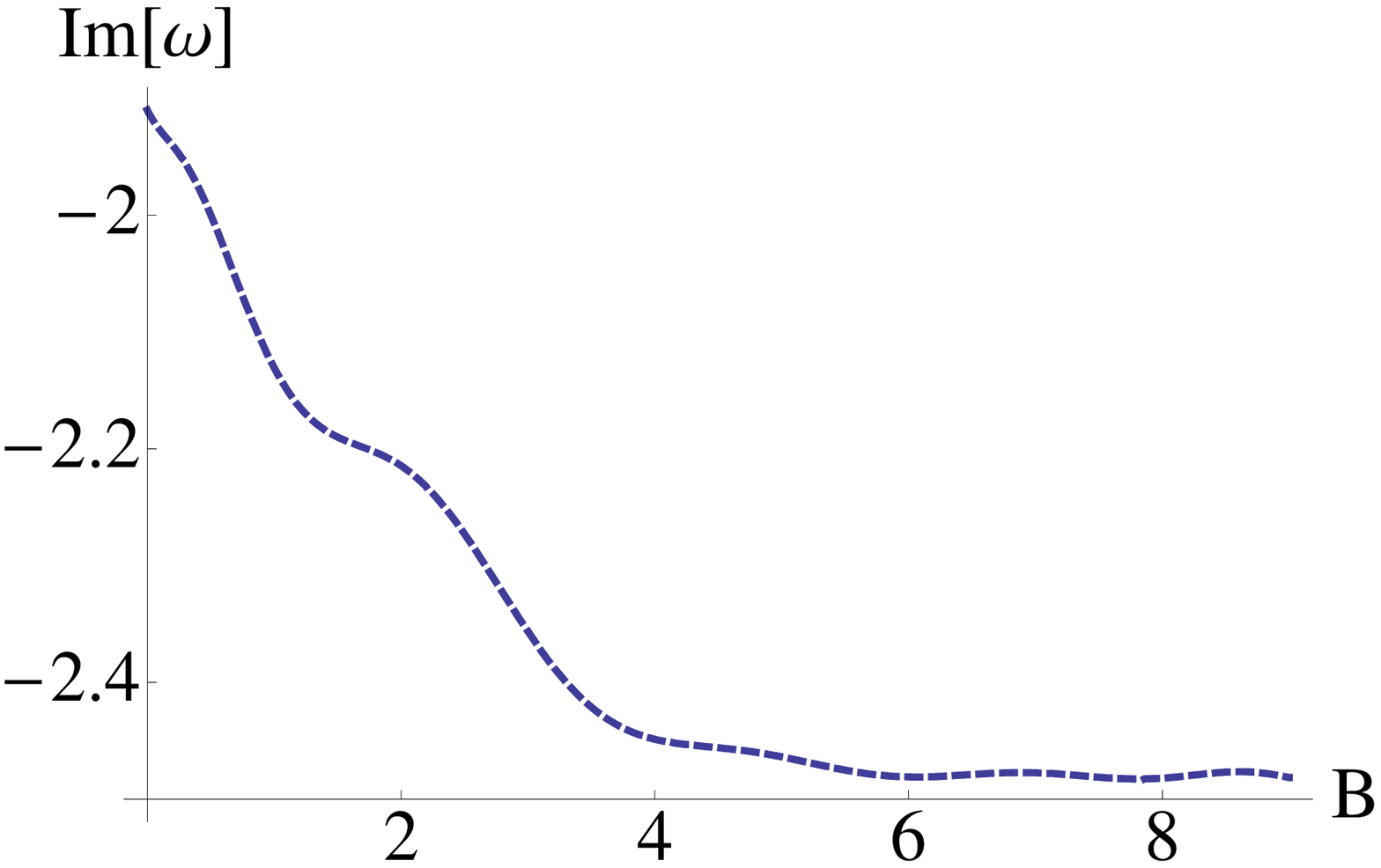} \label{fig: sI}}
\caption{\small The real ($\omega_{\rm R}={\rm Re}[\omega]$) and imaginary ($\omega_{\rm I}={\rm Im}[\omega]$) part of the quasinormal modes for the scalar fluctuation as a function of the external field. The quasinormal frequency is measured in units of the background temperature $T$. The small wiggles are due to numerical errors; here we have plotted the polynomial which fits the data. As expected before, we confirm that ${\rm Abs}[\omega_{\rm R}]> {\rm Abs}[\omega_{\rm I}]$, however they are roughly of the same order.}
\end{center}
\end{figure}

Let us now represent the quasinormal frequency as $\omega=\omega_{\rm R}+i\omega_{\rm I}$. It can be noted from the effective Schr\"{o}dinger equation that there is a $\mathbb{Z}_2$ symmetry in $\omega_{\rm R}$ (namely, if $g(u)$ is a solution with a frequency $\omega$ then $g^*(u)$ is also a solution with frequency $-\omega^*$\cite{Hoyos:2006gb}). Now we can observe that the Schr\"{o}dinger potential increases monotonically as we move away from the horizon. This implies that ${\rm Re}[\omega^2]>0$ and hence ${\rm Abs}[\omega_{\rm R}]> {\rm Abs}[\omega_{\rm I}]$. By explicit numerical calculation we will find that this is indeed the case (we point to figure \ref{fig: sR} and \ref{fig: sI}).

To extract more of the physics, we now compute the quasinormal frequency corresponding to this scalar fluctuation. We find it extremely convenient to perform a change of coordinates as implemented in ref.~\cite{Evans:2008tv}. Let us introduce a new variable $\rho=u/U_T$ and $v=t+\alpha(\rho)$, where $\alpha(\rho)$ is determined from the condition that 
\begin{eqnarray}\label{eqt: vt}
dv=dt+\frac{1}{\rho^3f(\rho)}d\rho\ .
\end{eqnarray}
This coordinate system $\{v,\rho\}$ makes the numerics conveniently stable\footnote{We refer the reader to ref. \cite{Evans:2008tv} for more details.} to find the quasinormal modes. The equation of motion\footnote{In presence of an external field this equation does not take the general form described in ref. \cite{Evans:2008tv}. This is due to the breaking of the $SO(3,1)\to SO(1,1)\times SO(2)$ by the presence of the magnetic field as a result of which the Laplacian in the $\{x^2, x^3\}$-direction gets squashed.} for the fluctuation now takes the form
\begin{eqnarray}
\partial_\rho\left(\rho^{11/2}(1+B^2\rho^{-3})^{1/2}ff_{\chi}'\right)-i\omega\left[\partial_\rho\rho^{5/2}(1+B^2\rho^{-3})^{1/2}f_{\chi}+2\rho^{5/2}(1+B^2\rho^{-3})^{1/2}f_{\chi}'\right]\nonumber\\
+\rho^{5/2}(1+B^2\rho^{-3})^{1/2}f_{\chi}=0\ .
\end{eqnarray}
It is now straightforward to check that this equation admits the existence of quasinormal modes. To find how it depends on the background magnetic field we use a shooting technique to solve the fluctuation equation and pick the appropriate value of $\omega$ for which the equation admits a normalizable solution (this is achieved by imposing the condition that $f_{\chi}\to 0$ as $\rho\to\infty$). As boundary conditions we impose $f_{\chi}(1+\epsilon)=1$ and $f_{\chi}'(1+\epsilon)$ equal to a value obtained by requiring the equation of motion to be regular near $\rho=1$, the event horizon. Here the parameter $\epsilon$ is a vanishingly small number in our numerical scheme. 

\label{fig: can}

The result of this pursuit is summarised in figure \ref{fig: sR} and \ref{fig: sI}. Since we are in the high temperature phase the quasinormal mode is comprised of both real and imaginary parts. It would be expected that the absolute value of the real part corresponds to the mass of the meson before it melts and the absolute value of the imaginary part corresponds to its inverse lifetime analogous to ref.~\cite{Hoyos:2006gb} for the D3--D7 system. Figure \ref{fig: sR} then tells us that the magnetic field increases the scalar meson mass and figure \ref{fig: sI} hints that the inverse lifetime gets bigger as the field is dialed up. However the precise identification of the meson mass and inverse lifetime with the quasinormal modes of the scalar fluctuation is a subtle issue in this particular model (see ref.~\cite{Paredes:2008nf}).

\subsection{Vector Fluctuation}

The vector meson spectrum in general is determined by a set of coupled equations of motion. It is however possible to completely decouple the transverse and the longitudinal modes if we restrict ourselves to the oscillations parallel to the magnetic field. If we restrict ourselves to the oscillation in one of the perpendicular directions (e.g. along the $x^3$-direction) to the applied magnetic field, then it is also possible to decouple one of the transverse modes (in this case the $A_2$-vector mode). However, the remaining transverse mode and the longitudinal mode remain coupled. The details are given in appendix A. Here we study the decoupled sectors only.

\subsubsection{The Transverse Mode}

We begin analyzing the transverse vector meson spctra. To extract qualitative features of the spectra we study the Schr\"{o}dinger potential. The equation of motion for the $A_2$ vector modes is given by
\begin{eqnarray}
\partial_u\left[e^{-\phi}\sqrt {-\det \left(E_{ab}^{(0)}\right)} \mathcal{S}^{22}\mathcal{S}^{uu} A_2'(u)\right]+\left(e^{-\phi}\sqrt {-\det \left(E_{ab}^{(0)}\right)}\right)\omega^2\mathcal{S}^{tt}\mathcal{S}^{22}A_2(u)=0\ ,
\end{eqnarray}
where we chose an ansatz of the form $A_2=A_2(u){\rm exp}(-i\omega t)$. It is straightforward to turn this equation in the form of a Schr\"{o}dinger equation with an effective potential given by
\begin{eqnarray}
&& V_S(u)=\frac{\left(u^3-1\right) \left(5 \left(7 u^3+5\right) B^4-2 \left(u^6-43 u^3\right) B^2+8 \left(u^9+2 u^6\right)\right)}{16 u^5\left(u^3+B^2\right)^2}\ , \nonumber\\
&& {\rm with} \quad d\tilde{u}=\frac{u^{3/2}}{u^3-1}du\ ,
\end{eqnarray}
where $\tilde{u}$ is the ``tortoise" coordinate. Clearly the horizon is located at $\tilde{u}\to\infty$ and the boundary is located at $\tilde{u}\to 0$.

\begin{figure}[!ht]
\begin{center}
\includegraphics[angle=0,
width=0.55\textwidth]{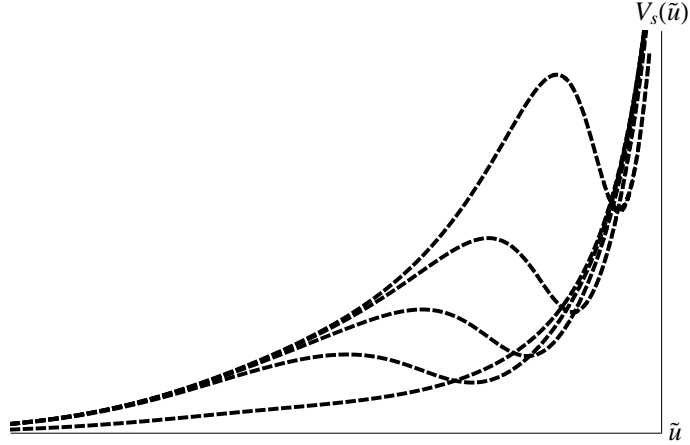}
\caption{\small The Schr\"{o}dinger potential for the vector meson for different values of external magnetic field. Here $\tilde{u}$ denotes the ``tortoise" coordinate mentioned in appendix B. As the magnetic field increases we observe that a small well develops in the effective Schr\"{o}dinger potential. The depth of this positive well is set by the background field.}
\label{fig: A2sch}
\end{center}
\end{figure}

The corresponding potential is shown in figure \ref{fig: A2sch}. The main qualitative feature that emerges from this is the fact that there is no instability due to existence of bound states of positive imaginary frequency modes since the potential remains positive definite. We observe, however, that a non-vanishing magnetic field can create a dimple in the otherwise smooth and rather featureless potential and therefore create local (positive) potential well. Therefore it is possible to have long-lived bound states with ${\rm Abs}[\omega_{\rm R}]\gg {\rm Abs}[\omega_{\rm I}]$. In this case, ${\rm Abs}[\omega_{\rm R}]$ is set by the depth of the well, which in turn is set by the background field. If we increase the magnetic field even higher then the dimple gets bigger and bigger and ultimately swallows the well. At even higher magnetic field the potential again takes a shape similar to the vanishing external field case. Therefore the imaginary part of the quasinormal frequency may have an interesting behaviour with increasing magnetic field. We may expect that  ${\rm Abs}[\omega_{\rm I}]$ starts decreasing with increasing magnetic field, but after a critical value it starts increasing again. Next we turn to the numerics.

The equation of motion (for the vector mode $A_2$ with non-zero spatial momentum) in this case is given by
\begin{eqnarray}
\partial_a\left[e^{-\phi}\sqrt {-\det \left(E_{ab}^{(0)}\right)} S^{22}\left(S^{a\rho}F_{\rho2}+S^{av}F_{v2}+S^{ai}F_{i2}\right)\right]=0\ , \quad i\in \{1,3\}\ ,
\end{eqnarray}
where $S^{ab}$ are the analogue of $\mathcal{S}^{ab}$ in eqn. (\ref{eqt: comp}) when the coordinate change described in eqn.~(\ref{eqt: vt}) has been performed. 

The background field breaks the full ${\rm SO}(3,1)$ Lorentz symmetry and there is an unbroken ${\rm SO}(2)$ symmetry corresponding to the rotation in the plane perpendicular to the magnetic field. Therefore the transverse vector meson spectra with a non-zero momentum will have two distinct branches. Mesons can have momentum along the direction of the background field (corresponding to $i=1$) or in the perpendicular plane (corresponding to $i=3$).  

Without any loss of generality we can consider studying the spectrum of the $A_2$ vector meson. We can give this meson a momentum along the $x^1$-direction or along the $x^3$-direction (but no momentum along $x^2$-direction since this is the transverse mode). The spectrum would be entirely equivalent to the spectrum of the $A_3$ vector modes having momentum along the $x^1$-direction or along the $x^2$-direction.

Now we follow the same numerical approach. We impose $A_2(1+\epsilon)=1$ and fix $A_2'(1+\epsilon)$ from the equation of motion near the horizon. With these boundary conditions we look for normalizable solutions for $A_2$, which gives the quasinormal modes.

\begin{figure}[!ht]
\begin{center}
\subfigure[] {\includegraphics[angle=0,
width=0.45\textwidth]{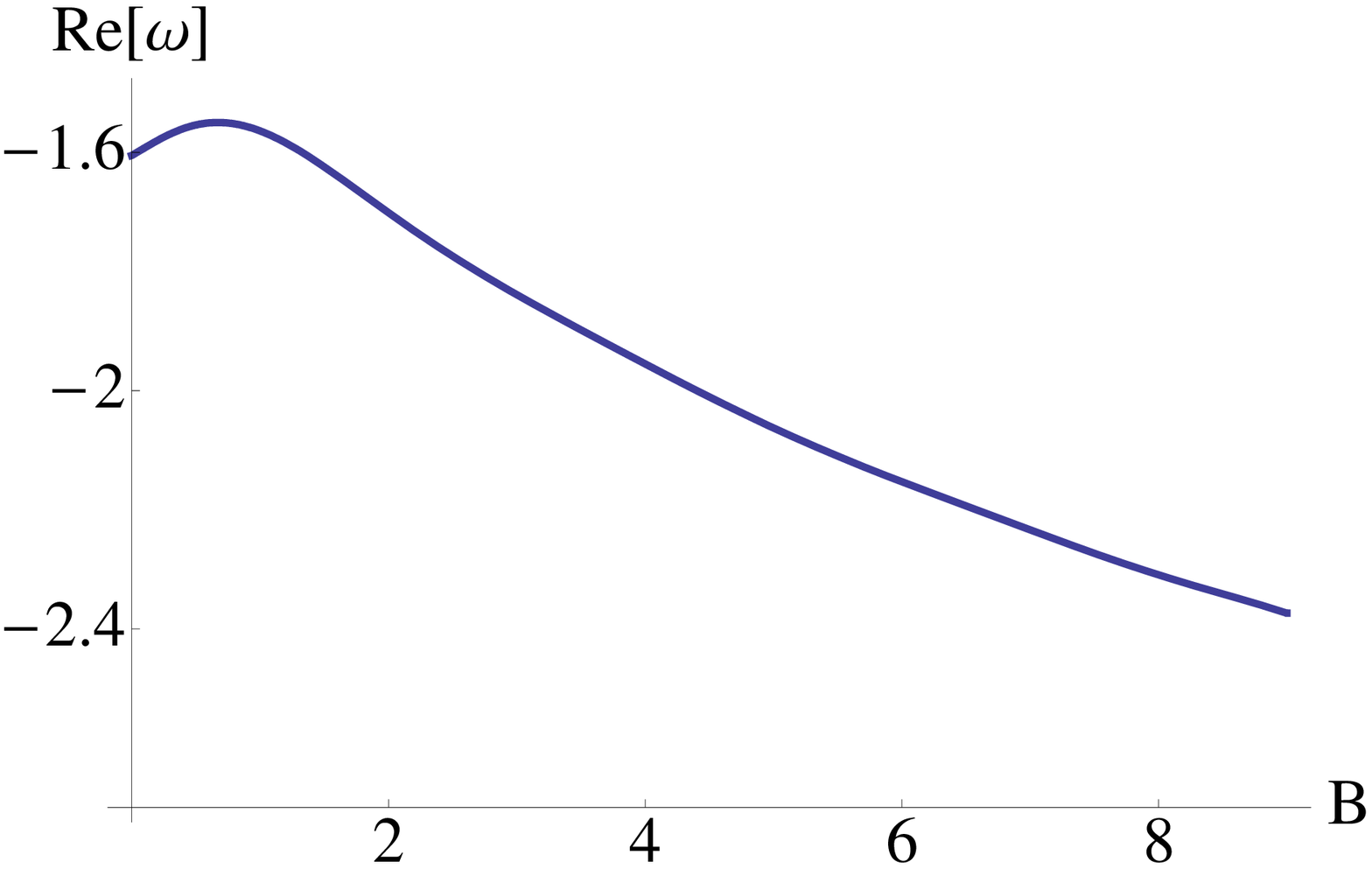} \label{fig: reb}}
\subfigure[] {\includegraphics[angle=0,
width=0.45\textwidth]{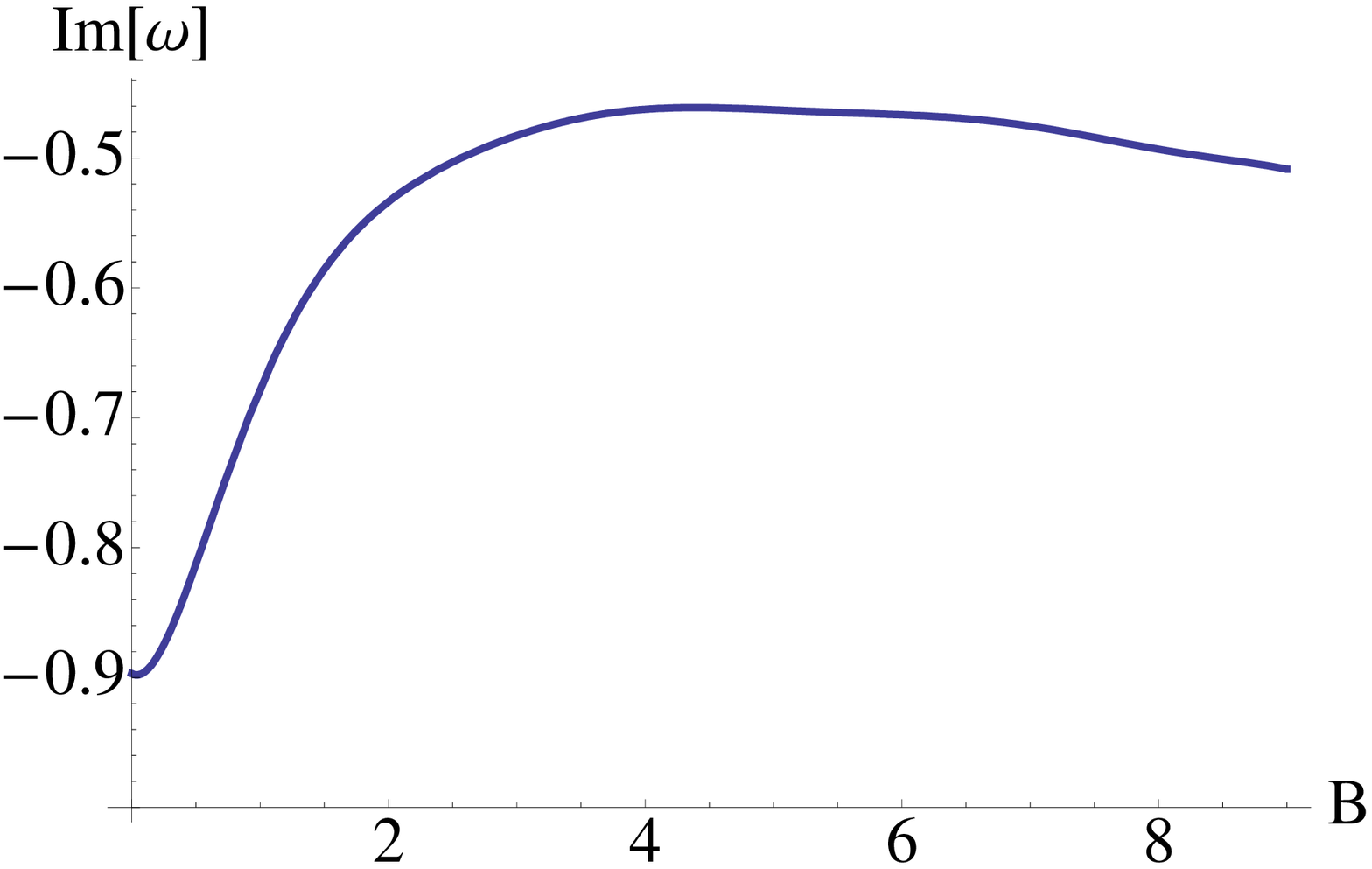} \label{fig: imb}}
\caption{\small The real and imaginary part of the quasinormal modes for the vector fluctuation as a function of the external field. The frequency is expressed in units of background temperature $T$ and the magnetic field is expressed in units of $R$. We find a local minima in ${\rm Abs}[\omega_{\rm I}]$ in the parameter range we have explored. This is qualitatively similar to the behaviour anticipated from analyzing the effective Schr\"{o}dinger potential. We also observe that in the most of the parameter space ${\rm Abs}[\omega_{\rm R}]\gg {\rm Abs}[\omega_{\rm I}]$.}
\end{center}
\end{figure}

\begin{figure}[!ht]
\begin{center}
\subfigure[] {\includegraphics[angle=0,
width=0.45\textwidth]{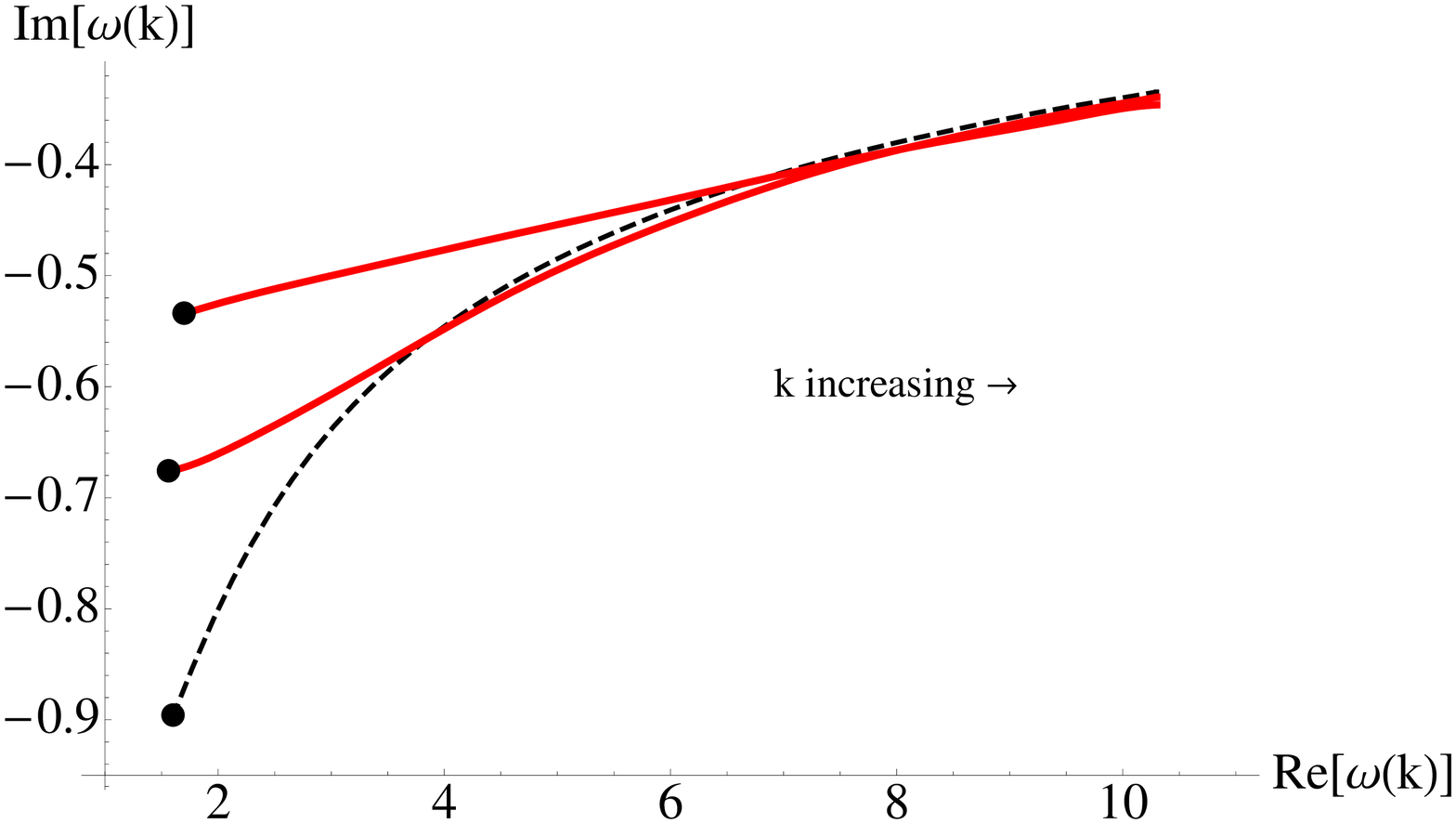} \label{fig: dis}}
\subfigure[] {\includegraphics[angle=0,
width=0.45\textwidth]{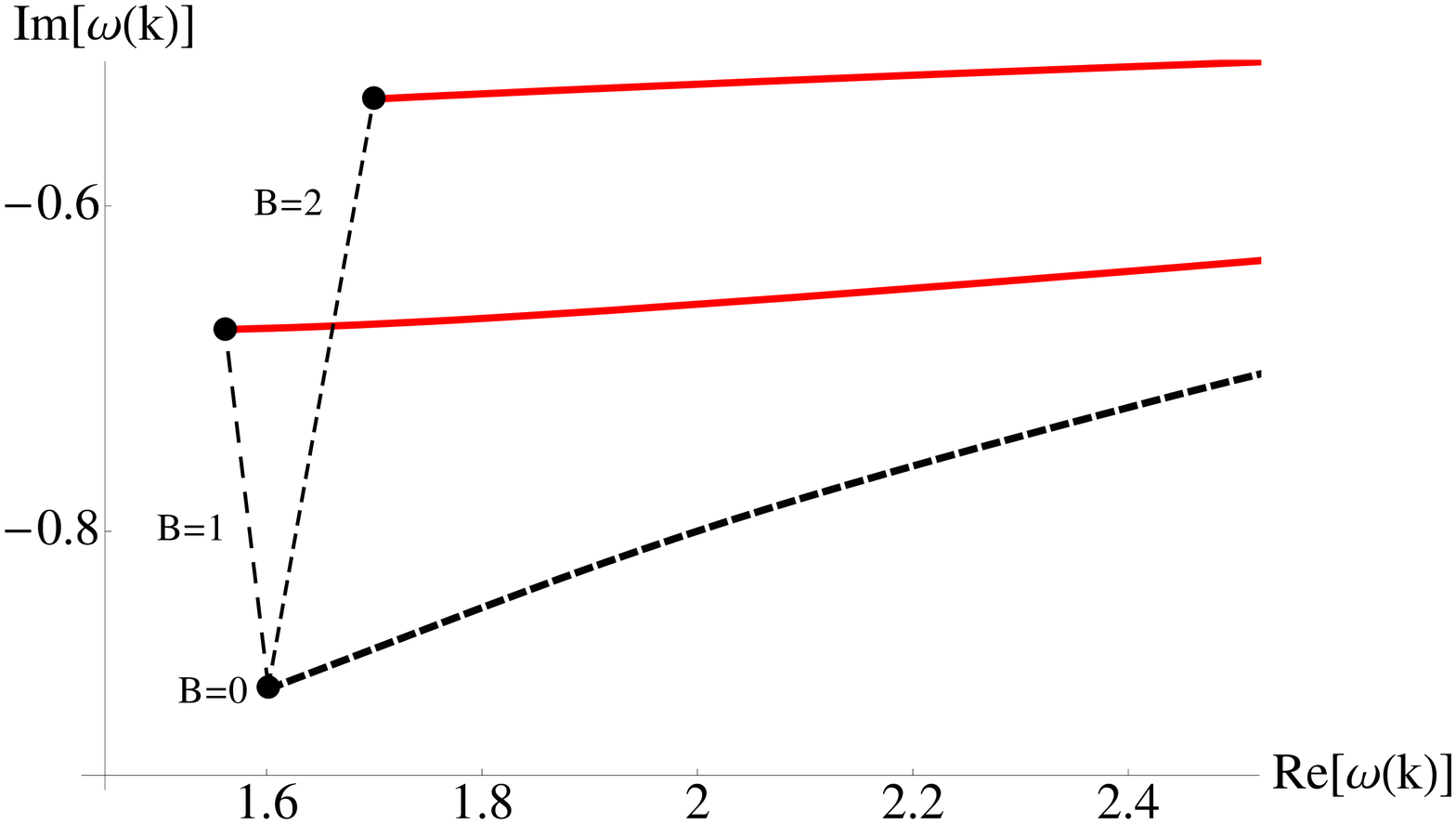} \label{fig: dissplit}}
\subfigure[]{\includegraphics[angle=0,
width=0.45\textwidth]{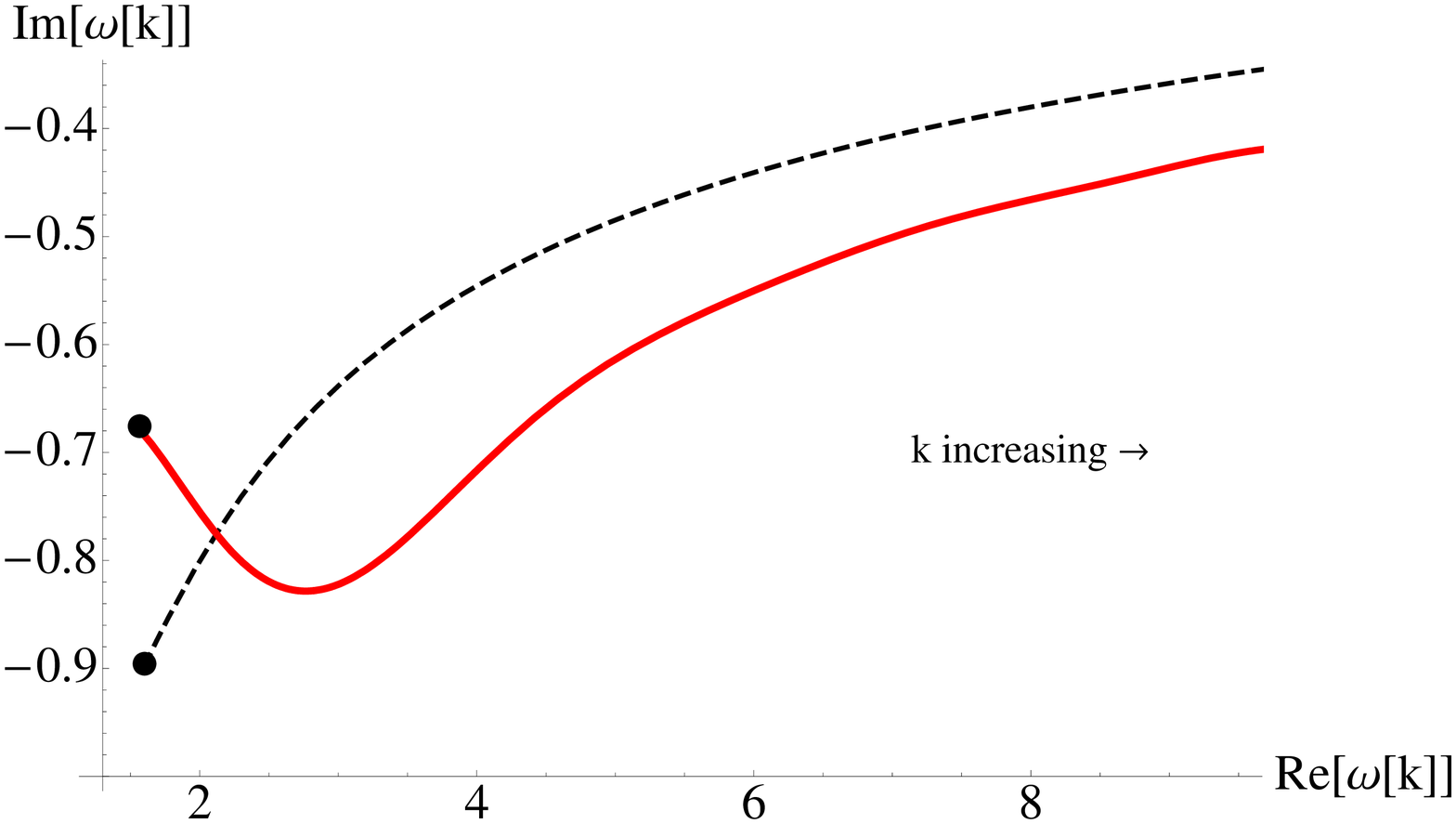} \label{fig: dis3}}
\caption{\small The dispersion relation between the real and imaginary part of the quasinormal frequency for two different values of magnetic field. The red (solid) curves correspond to the dispersion relation for non-zero magnetic field. The black (dashed) curve corresponds to the dispersion relation when there is no magnetic field. The black dots represent the corresponding values of quasinormal frequency for vanishing spatial momentum. Figure \ref{fig: dis} and \ref{fig: dissplit} correspond to vector mesons having momenta along $x^1$-direction and figure \ref{fig: dis3} corresponds to vector meson with momentum along $x^3$-direction.}
\end{center}
\end{figure}

In figure \ref{fig: reb} and \ref{fig: imb} we have shown the dependence of the real and imaginary part of quasinormal frequency for the particular choice of $k=0$. We observe that the inverse lifetime of the vector meson gets shorter as the external field is dialed up and therefore enhances the stability of the meson. This behaviour is opposite to what we observed for the scalar meson fluctuation.

Next we obtain the dispersion relation when $k\not=0$. In figure \ref{fig: dis} we demonstrate this dispersion relation for a wide range of values for $k$ and in figure \ref{fig: dissplit} we have shown a magnified version of figure \ref{fig: dis} to more clearly see the role an external magnetic field plays in this case. These two figures correspond to the dispersion relation when the momentum is parallel (i.e. along $x^1$-direction) to the magnetic field. We observe that for high enough spatial momentum the dispersion relation curve approximates the zero field dispersion curve quite well. We have shown in figure \ref{fig: dis3} the dispersion relation when the momentum is perpendicular (i.e. along $x^3$-direction) to the background field. The qualitative behaviour between these two dispersion curves are clearly different; however for large enough spatial momenta these curves tend to become insensitive to the background field. This is because the heavier the meson becomes the less sensitive  it is to the background field.

\subsubsection{The Longitudinal Modes}

Now we study the longitudinal mode. With the change in variables in equation (\ref{eqt: vt}), we write down the equations of motion for gauge fields along  $\{v, x^1\}$
\begin{eqnarray}
&& \partial_{\rho}\left[e^{-\phi}\sqrt{-{\rm det}\left(E_{ab}^{(0)}\right)}\left(S^{vv}S^{\rho\rho}-S^{v\rho}S^{\rho v}\right)A_v'\right]\nonumber\\
&& - e^{-\phi}\sqrt{-{\rm det}\left(E_{ab}^{(0)}\right)}k\left[S^{vv}(kA_v+\omega A_1)+i  S^{v\rho} A_1'\right] + 4 g_s^{-1} B A_1'=0\ ,\nonumber
\end{eqnarray}
\begin{eqnarray}
&& \partial_{\rho}\left[e^{-\phi}\sqrt{-{\rm det}\left(E_{ab}^{(0)}\right)}S^{11}\left(S^{\rho\rho} A_1'-i S^{\rho v}(kA_v+\omega A_1)\right)\right]\nonumber\\
&& - e^{-\phi}\sqrt{-{\rm det}\left(E_{ab}^{(0)}\right)}S^{11}\omega\left[S^{vv}(kA_v+\omega A_1)+i  S^{v\rho} A_1'\right] - 4 g_s^{-1} B A_v'=0\ ,
\end{eqnarray}
along with the constraint
\begin{eqnarray}
&& e^{-\phi}\sqrt{-{\rm det}\left(E_{ab}^{(0)}\right)}\left(S^{\rho\rho}S^{vv}-S^{\rho v}S^{v\rho}\right)i\omega A_v' + i4 g_s^{-1} B  (\omega A_1+k A_v)\nonumber\\
&& -q e^{-\phi}\sqrt{-{\rm det}\left(E_{ab}^{(0)}\right)}S^{11}\left[i S^{\rho\rho}A_1'+S^{\rho v}(k A_v+\omega A_1)\right]=0\ ,
\end{eqnarray}
where $k$ is the momentum along $x^1$-direction. We define the longitudinal mode to be given by
\begin{equation}
\omega A_1+k A_v= \mathbb{E}\ .
\end{equation}
Now using the definition of the longitudinal mode and the constraint equation we can solve for $A_v'$ and $A_{\rho}'$ in terms of $\mathbb{E}$ and $\mathbb{E}'$ and then substitute back in the equation of motion for the gauge fields to obtain
\begin{eqnarray}\label{eqt: longEv}
\mathbb{E}''(\rho)+\mathcal{Z}_1(\rho)\mathbb{E}'(\rho)+\mathcal{Z}_2(\rho)\mathbb{E}(\rho)=0\ ,
\end{eqnarray}
where $\mathcal{Z}_1$ and $\mathcal{Z}_2$ are known functions\footnote{The explicit expressions of $\mathcal{Z}_1$ and $\mathcal{Z}_2$ are not very illuminating, thus we do not provide them here.} of $\rho$.

Equation (\ref{eqt: longEv}) now can be numerically solved by using the boundary conditions that $\mathbb{E}(1+\epsilon)=1$ and $\mathbb{E}'(1+\epsilon)$ equal to a value obtained from the equation of motion itself. The condition of normalizability then fixes the quasinormal modes.

\begin{figure}[!ht]
\begin{center}
\subfigure[] {\includegraphics[angle=0,
width=0.45\textwidth]{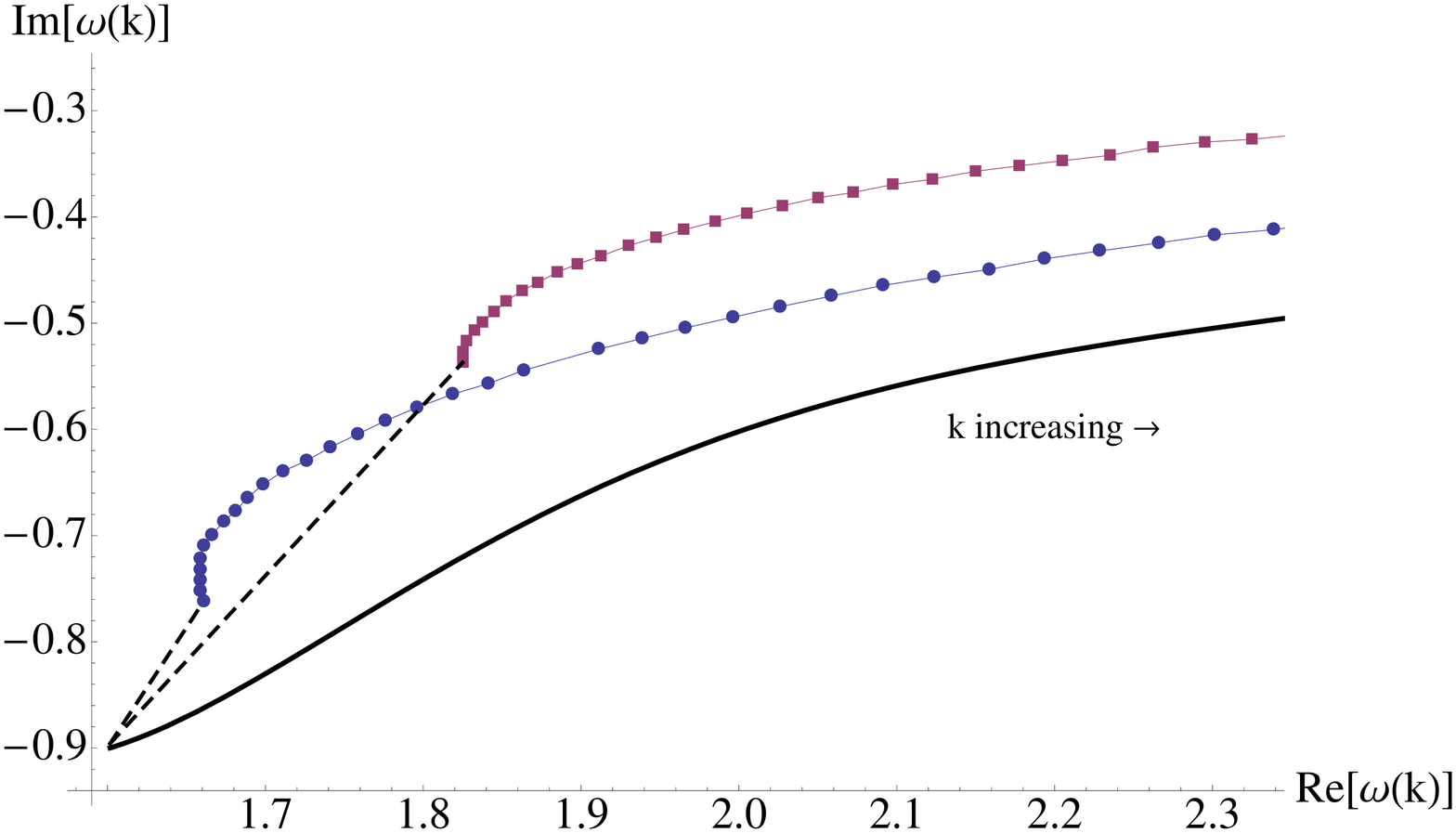} \label{fig: quasilong}}
\subfigure[] {\includegraphics[angle=0,
width=0.45\textwidth]{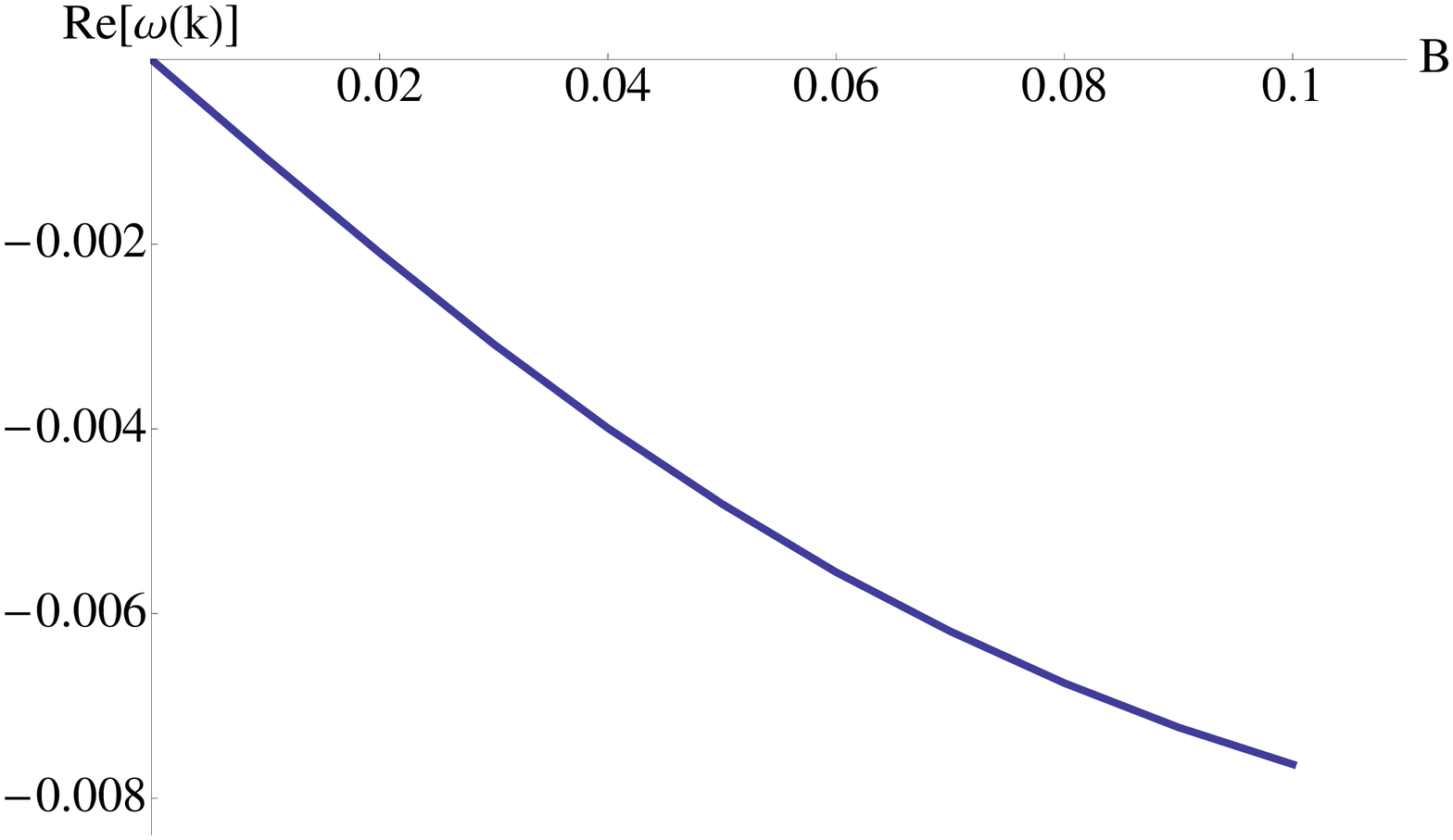} \label{fig: hydrolong}}
\caption{\small The dispersion relation of longitudinal oscillation in presence of a background magnetic field. The quasinormal frequency is measured in units of the background temperature. In figure \ref{fig: quasilong} the black solid curve represents the dispersion relation when the magnetic field is absent, the blue (circular) dots correspond to $B=5\times 10^{-2}$ and the purple (square) dots correspond to $B=10^{-1}$. $B$ has been expressed in units of $R$. In figure \ref{fig: hydrolong} we have shown the dependence of the lowest quasinormal frequency corresponding to the hydrodynamic mode with the magnetic field for a fixed momentum $k=0.01$. The imaginary part of the quasinormal frequency does not depend on the magnetic field in the hydrodynamic limit.}
\end{center}
\end{figure}

The numerical result is shown\footnote{Note that here we numerically explore only for positive values of $B$. The equation of motion for the longitudinal modes have a linear term in $B$, and therefore the quasinormal modes may depend on the sign of $B$. However, we found numerically that the qualitative features of the quasinormal modes are similar also for negative values of $B$.} in figure \ref{fig: quasilong}. The qualitative behaviour of the dispersion curve is similar to the transverse modes. It is evident that increasing magnetic field brings quasinormal frequency closer to the real axis in the complex plane. However, for a fixed magnetic field, the dispersion curves tend to approach the zero magnetic field dispersion curve as the momentum increases.

\section{Mesons with Large Spin}

Mesons with large spin can be described by strings having both its end points on the flavour brane and rotating on a two plane along the directions where the gauge theory lives. For large enough angular momentum, the length of the string is much larger compared to $\ell_s$ and we can use classical Nambu-Goto action to describe the dynamics of the rotating string. Mesons with large spin have previously been analyzed in, {\it e.g.}, refs. \cite{Peeters:2006iu, Kruczenski:2003be}. Here we will follow the same framework, now including the effect of an external magnetic field.

In absence of magnetic field due to rotational symmetry, the spectrum does not depend on the direction of the angular momentum $J$. An external magnetic field breaks this symmetry and introduces a splitting of the energy levels (Zeeman effect). Here we will focus on the two cases, namely when the magnetic field is parallel or anti-parallel to the angular momentum. In this case there should be a Zeeman splitting in energy for mesons with non-zero magnetic moment.

This idea has been implemented in ref.~\cite{Jensen:2008yp} in a different model\footnote{For more recent similar work see, {\it e.g.}, refs.~\cite{Haque:2009hz, Colgain:2009zh}.}. We adopt similar system and consider two flavour brane--anti-brane pair very close to each other so that in the bulk they join leaving two diagonal $U(1)$s (call this $U(1)_A$ and $U(1)_B$ corresponding to the two pair of flavour branes). We consider now turning on fields having equal opposite charge under these two $U(1)$s. Now $AA$ and $BB$ strings will have equal and opposite charges but $AB$ and $BA$ strings will have equal charges. In the first case, the total orbital magnetic moment of the meson will vanish whereas in the second case the meson will have a non-zero magnetic moment. The magnetic field will induce a quadratic correction to the meson energy for the former and a linear Zeeman splitting for the later. Here we will analyze the later case only. It is useful to note that this configuration is symmetric under reflection around the midpoint of the string (at $\rho=0$). We will refer to this as the symmetric configuration.

To proceed we write the relevant part of the background metric\footnote{Here we explicitly write down the metric corresponding to the high temperature phase only. The background metric corresponding to the low temperature phase can likewise be written in an analogous form.} in the following form (rewriting the $\{x^2,x^3\}$-two plane in polar coordinate $\{\rho,\phi\}$)
\begin{eqnarray}
ds^2&=&\left(\frac{u}{R}\right)^{3/2}\left(f(u)dt_E^2+d\rho^2+\rho^2d\phi^2\right)+\left(\frac{R}{u}\right)^{3/2}\frac{du^2}{f(u)}\ ,\nonumber\\
f(u)&=&1-\left(\frac{u_T}{u}\right)^3\ .
\end{eqnarray}
We recover the zero temperature background by setting $u_T=0$. We will work with the Nambu-Goto action for the string with the following ansatz for the string profile
\begin{eqnarray}
t=\tau\ , \quad \rho=\rho(\sigma)\ , \quad u=u(\sigma)\ , \quad \phi=\omega\tau\ .
\end{eqnarray}
From now on the parameters $\{\tau,\sigma\}$ will always refer to the worldsheet coordinates for the string. We also assume that $\omega$ is positive (i.e. only clockwise rotation).

In the presence of an external constant magnetic field the Nambu-Goto action is accompanied by a boundary term coupled to the magnetic potential $A_{\phi}=B\rho^2/2$. This potential $A_{\phi}$ gives rise to a magnetic field $B_{(2)}=B\rho d\rho\wedge d\phi=Bdx^2\wedge dx^3$. Taking this boundary term into account the action for the string is given by
\begin{eqnarray}\label{eqt: action}
&& S = \frac{1}{2\pi\alpha'}\int d\tau d\sigma \left[\left(\frac{u}{R}\right)^3\left(f(u)-\rho^2\omega^2\right)\left(\rho'^2+\frac{u'^2}{f(u)}\left(\frac{R}{u}\right)^3\right)\right]^{1/2}+\Delta S_{B}\ ,\nonumber\\
&& \Delta S_B = \frac{1}{2\pi\alpha'}\left[\int \left. A\right |_{\sigma^+} + \int \left. A\right |_{\sigma^-}\right]\ , \quad A=\frac{B\rho^2}{2}d\phi\ ,
\end{eqnarray}
where $\sigma^{\pm}$ represents the right and the left boundaries and the relative positive sign between the two boundary contributions coming from the magnetic potential is due to the  symmetric configuration\footnote{For the non-symmetric configuration the there will be a relative negative sign between these two terms. We refer to ref. \cite{Jensen:2008yp} for further details on this construction.}.
For convenience we introduce the rescaled variables
\begin{equation}
\tilde{\rho}=\rho\omega\ , \quad \tilde{u}=u\omega\ , \quad \tilde{R}=R\omega\ , \quad \tilde{\sigma}=\sigma\omega\ .
\end{equation}
The expression for the energy and the angular momentum for the spinning string in the rescaled variables can simply be obtained to give
\begin{eqnarray}\label{eqt: ej}
&& E = \omega\frac{\partial L}{\partial\omega}-L=\frac{1}{2\pi\alpha'}\left(\frac{1}{\omega^{5/2}R^{3/2}}\right)\int d\tilde{\sigma}\tilde{u}^{3/2}\left(\tilde{\rho}'^2+\tilde{u}'^2\left(\frac{\tilde{R}}{\tilde{u}}\right)^3\right)^{1/2}\frac{1}{\sqrt{1-\tilde{\rho}^2}}\ , \nonumber\\
&& J = \frac{\partial L}{\partial\omega}=\frac{1}{2\pi\alpha'}\left(\frac{1}{\omega^{7/2}R^{3/2}}\right)\int d\tilde{\sigma}\frac{\tilde{\rho}^2}{\sqrt{1-\tilde{\rho}^2}}\left[\tilde{u}^3\left(\tilde{\rho}'^2+\tilde{u}'^2\left(\frac{\tilde{R}}{\tilde{u}}\right)^3\right)\right]^{1/2}+\Delta J_B\ ,\nonumber\\
&& \Delta J_B=\frac{1}{2\pi\alpha'\omega^2}\left(\left. B\frac{\tilde{\rho}^2}{2}\right | _{\sigma^+} + \left. B\frac{\tilde{\rho}^2}{2}\right | _{\sigma^-}\right)\ .
\end{eqnarray}
We observe that the external field can enhance angular momentum of the meson if it is aligned in parallel (positive values of $B$); and can reduce the angular momentum if it is anti-parallel (negative values of $B$).

To proceed we need to solve for the string embeddings. The equation of motion obtained from the Nambu-Goto action has to be supplemented by the boundary condition
\begin{eqnarray}
\left. \pi_X^1\right |_{\partial\Sigma}=\left. \frac{\partial L}{\partial (X')^M}\delta X^M \right |_{\partial\Sigma}=0\ .
\end{eqnarray}
Since $\left. \delta u\right |_{\partial\Sigma}=0$ and $\left. \delta\rho\right |_{\partial\Sigma}$ is arbitrary, we have to impose the Neumann boundary condition (in absence of any external field) $\left. (\partial L/\partial \rho')\right |_{\partial\Sigma}=0$. This condition gives us the following constraint
\begin{eqnarray}\label{eqt: bc}
\left. \pi_{\tilde{\rho}}^1\right |_{\partial\Sigma}=\left. \tilde{\rho}'\left[\frac{\left(\tilde{u}/R\omega\right)^3(1-\tilde{\rho}^2)}{\tilde{\rho}'^2+\tilde{u}'^2\left(\omega R/\tilde{u}\right)}\right]^{1/2}\right |_{\partial\Sigma}=0\ .
\end{eqnarray}
So we need to impose the boundary condition $\left. \tilde{\rho}'\right |_{\partial\Sigma}=0$, which means that the string is ending perpendicularly on the flavour brane--anti-brane pair. We will come back to the details of our numerical scheme in a later section.

The presence of an external magnetic field changes this boundary condition. It can be easily seen that the boundary condition to be satisfied is now given by
\begin{eqnarray}\label{eqt: bcb}
\left. \pi_{\tilde{\rho}}^1\right |_{\partial\Sigma}-\left(-B\tilde{\rho}\right)=0\ ,
\end{eqnarray}
where $\left. \pi_{\tilde{\rho}}^1\right |_{\partial\Sigma}$ is given in equation (\ref{eqt: bc}). Therefore we need to solve the equation of motion obtained from the action in eqn. (\ref{eqt: action}) subject to the boundary condition given in eqn. (\ref{eqt: bc}) or (\ref{eqt: bcb}) in absence or presence of a magnetic field respectively.

We will begin by obtaining some analytical results in the large angular frequency limit in the zero temperature phase. We will choose $\sigma=\rho$ gauge and send $\omega\to\infty$; from the reality condition of the Nambu-Goto action in eqn. (\ref{eqt: action}) this limit sets a bound $|\rho|<1/\omega\to 0$; and therefore equivalently we study the short string limit.

\subsection{Analytical Results}

In this section we consider the zero temperature background and spinning strings with large angular frequency only.

\subsubsection{Vanishing Magnetic Field}

In this section we closely follow the approach adopted in ref.~\cite{Kruczenski:2003be}. In the large angular frequency limit the relevant string profile is well approximated by the local geometry very close to the point from where the string hangs. In this limit the string hangs from the point where the brane--anti-brane pair join (the radial position denoted as $U_0$). We therefore pick the following ansatz to approximate the string profile
\begin{eqnarray}\label{eqt: oansat}
\tilde{u}(\tilde{\rho})=\omega U_0+\frac{f(\tilde{\rho})}{\omega}\ .
\end{eqnarray}
It is easy to check that this ansatz is consistent with the equation of motion for the string in $\omega\to\infty$ limit. To obtain the leading order behaviour we have to solve for the function $f(\tilde{\rho})$ with the boundary conditions that $f(0)=0=f'(0)$\footnote{At $\tilde{u}=\omega U_0$ we require $\tilde{u}'(\tilde{\rho})=0$, which sets the boundary condition for the function $f(\tilde{\rho})$.}. Substituting the ansatz given in eqn.~(\ref{eqt: oansat}) in the equation of motion for the string and keeping only the leading order terms we obtain a differential equation for the function $f(\tilde{\rho})$ which can be integrated to obtain the following analytic form
\begin{eqnarray}\label{eqt: anprofile}
f(\tilde{\rho})=\frac{3U_0^2}{8 R^3}\left(\tilde{\rho}^2+\arcsin^2(\tilde{\rho})\right)\ .
\end{eqnarray}
Plugging back $f(\tilde{\rho})$ from eqn. (\ref{eqt: anprofile}) to eqn. (\ref{eqt: ej}) and simplifying we get the following relations
\begin{eqnarray}\label{eqt: approx}
E=\frac{\pi U_0^{3/2}\lambda}{2 R^{9/2}}\frac{1}{\omega}+\mathcal{O}(\omega^{-5/2})\ ,\quad
J = \frac{\pi U_0^{3/2}\lambda}{4 R^{9/2}}\frac{1}{\omega^2}+\mathcal{O}(\omega^{-3})\ , \quad
E=\frac{\sqrt{\pi}U_0^{3/4}}{R^{9/4}}\sqrt{\lambda J} \ .
\end{eqnarray}
From this we can see that $\omega\to\infty$ limit is equivalent to $J\ll \lambda$ limit. Also we find that the mesons with large angular frequency follow a Regge trajectory with an effective tension\footnote{The radial position $U_0$ is related to the asymptotic separation $L$ between the flavours via $U_0^{1/2}=R^{3/2}/(4L) B(9/16,1/2)$, where $B(a,b)$ is the Beta function. This length $L$ sets the coupling strength of the non-local four Fermi interaction for the dual NJL model.} 
\begin{equation}\label{eqt: regge}
\tau_{\rm eff}=E^2/(2\pi J)=\lambda U_0^{3/2}/(2 R^{9/2}). 
\end{equation}

\subsubsection{Non-vanishing Magnetic Field}

\begin{figure}[!ht]
\begin{center}
\subfigure[] {\includegraphics[angle=0,
width=0.45\textwidth]{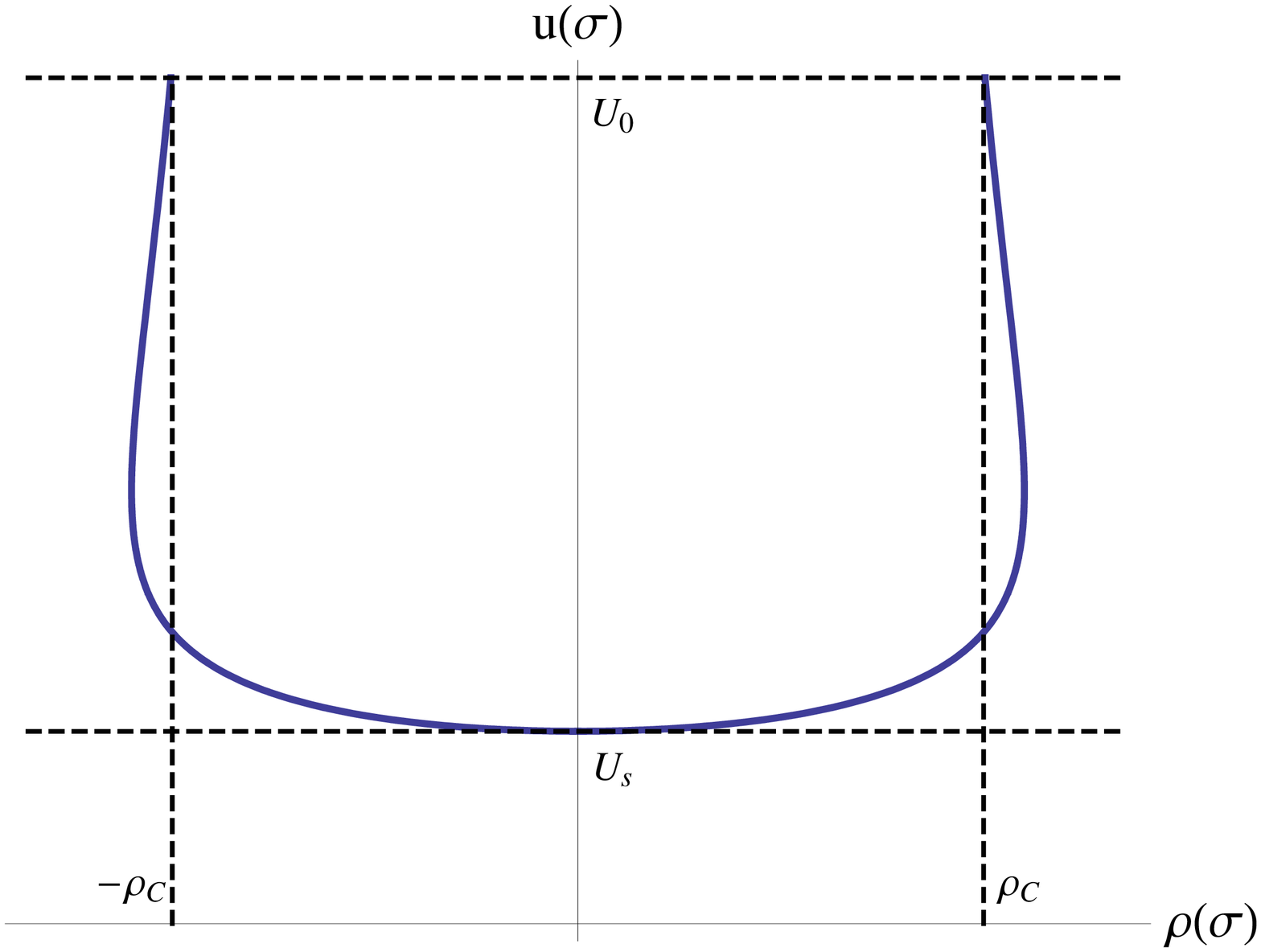} \label{fig: prohi}}
\subfigure[] {\includegraphics[angle=0,
width=0.45\textwidth]{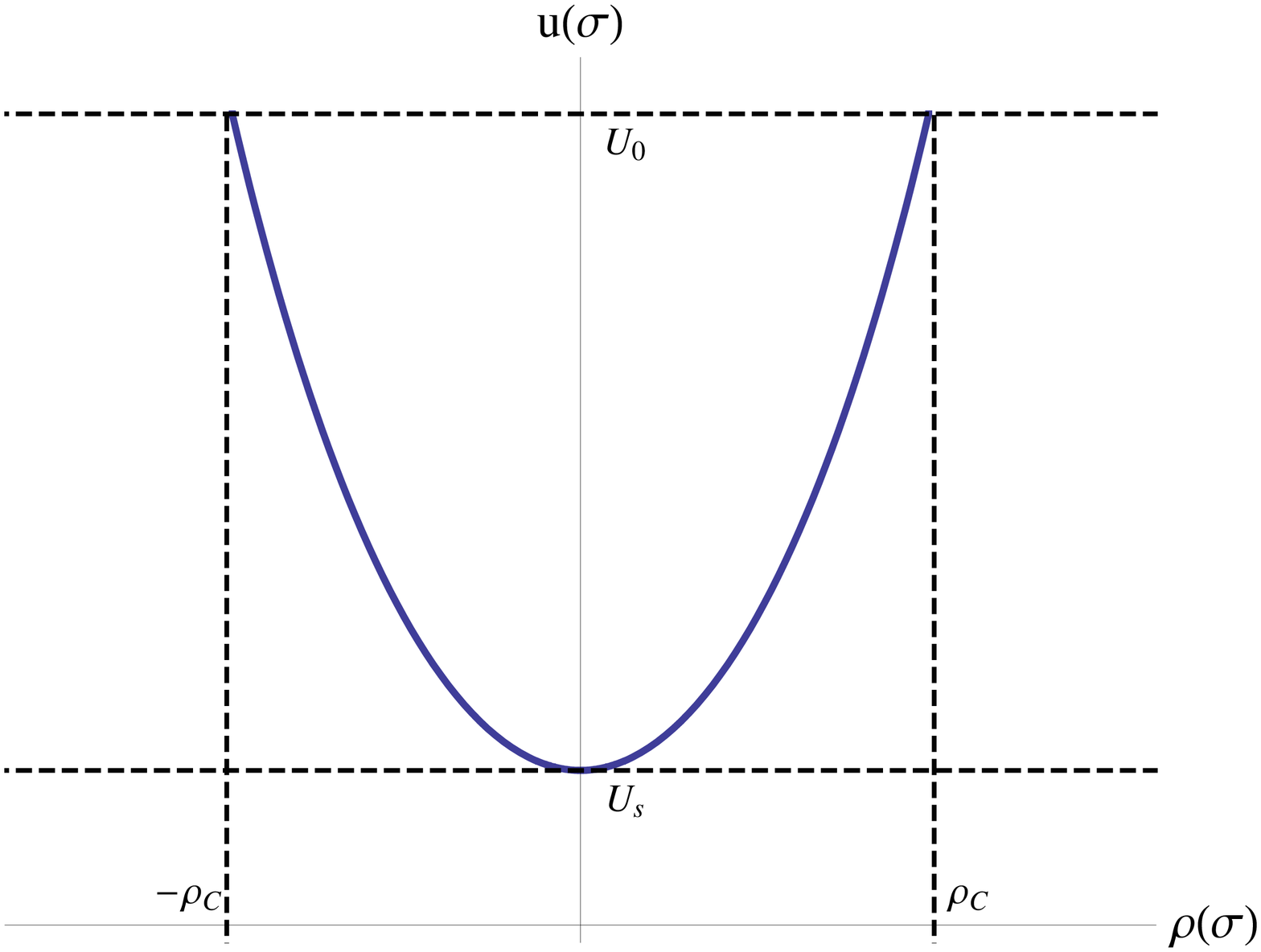} \label{fig: prolo}}
\caption{\small The two possible profiles are presented; $U_s$ is the radial turnaround position for the string. These are obtained by solving the equations of motion numerically, although we come back to the numerical results in a later section. The profile in the left hand side has more energy than the one in the right hand side.}
\end{center}
\end{figure}

In presence of a magnetic field the leading order solution for $\tilde{u}(\tilde{\rho})$ remains the same as in eqn. (\ref{eqt: anprofile}); however the boundary condition needs to be modified as given by eqn. (\ref{eqt: bcb}). Using the expression for $\left. \pi_{\tilde{\rho}}^1\right |_{\partial\Sigma}$ from eqn. (\ref{eqt: bc}) (in the gauge $\sigma=\tilde{\rho}$) in the modified boundary condition given in eqn. (\ref{eqt: bcb}) we can determine the value of $\tilde{\rho}$ at the boundary (meaning at $u=U_0$ where the string ends on the flavour brane). In the limit of small magnetic field this critical value of $\tilde{\rho}$ (denoted as $\tilde{\rho}_C$) can be obtained to be
\begin{eqnarray}
\tilde{\rho}_C=1-\frac{B^2}{2}\left(\frac{R}{U_0}\right)^3\ .
\end{eqnarray}
With this $\tilde{\rho}_C$, there will be two different string profiles which we have shown in figure \ref{fig: prohi} and \ref{fig: prolo} respectively.

These two different profiles result in two different energies (let us denote these energies by $E_{\pm}$). The one with the lower energy (i.e., with $E_{-}$) is extended from $\tilde{\rho}=0$ to $\tilde{\rho}_C$ and the one with the higher energy (i.e., with $E_{+}$) is extended from $\tilde{\rho}=0$ to $\tilde{\rho}=1$ and then folds back to $\tilde{\rho}=\tilde{\rho}_C$.
Using the formulae for meson energy $E$ and angular momentum $J$ given in eqn. (\ref{eqt: ej}) we get the following linear order corrections (Zeeman splitting) introduced by the magnetic field\footnote{The parameter $U_0$ is also related to the constituent quark mass. For a study of how the constituent quark mass behaves (for a fixed value of $L$) with the magnetic field see ref. \cite{Johnson:2008vna}. }
\begin{eqnarray}\label{eqt: splite}
E_{-} &=& \frac{\sqrt{\pi}U_0^{3/4}}{R^{9/4}}\sqrt{\lambda J}-\frac{2}{\sqrt{\pi}}\sqrt{\frac{J}{\lambda}}\frac{BR^{9/4}}{U_0^{3/4}}\ ,\nonumber\\
E_{+} &=& \frac{\sqrt{\pi}U_0^{3/4}}{R^{9/4}}\sqrt{\lambda J}+\frac{2}{\sqrt{\pi}}\sqrt{\frac{J}{\lambda}}\frac{BR^{9/4}}{U_0^{3/4}}\ .
\end{eqnarray}
It is easy to check that this splitting is consistent with the general formula for energy (obtained in ref.~\cite{Abouelsaood:1986gd}) of a Nambu-Goto string with an effective tension $\tau_{\rm eff}$ moving in a background magnetic field. The energy spectrum (for large $J$) is given by the following formula
\begin{eqnarray}\label{eqt: exact}
E^2=(2\pi\tau_{\rm eff})(1-\kappa) J\ , \quad \kappa=\frac{2}{\pi}\arctan\left(\frac{B}{\tau_{\rm eff}}\right)\ .
\end{eqnarray}
Recalling the expression for $\tau_{\rm eff}$ from eqn. (\ref{eqt: regge}) and using the general result above the linear order correction to the energy of the string can be obtained which matches exactly with what is obtained in eqn. (\ref{eqt: splite}).

\subsection{Numerical analysis}

In order to extract the relevant physics we need to solve for the string embedding. To that end we choose the parametrization $\sigma=\rho+u$ since it provides better numerical stability as discussed in {\it e.g.}, ref.~\cite{Jensen:2008yp}. 

To solve the equation of motion first we fix the maximum radial position the string can attain (this is the position where the brane--anti-brane pair join, which is denoted by $U_0$). This in turn fixes the asymptotic separation between the brane--anti-brane pair. Fixing $U_0$~is equivalent to fixing the constituent quark mass of the dual gauge theory and fixing $L$, the asymptotic separation of the brane--anti-brane pair, is equivalent to keeping the four fermi coupling of the dual NJL model. Now for a given $U_0$ we look for a radial turnaround position for the string (call it $U_s$) such that shooting from $U_s$ with the IR boundary conditions $\left. u(\sigma)\right |_{U_s}=U_s$ and $\left. u'(\sigma)\right |_{U_s}=0$ would satisfy the UV boundary condition $\left. u'(\sigma)\right |_{U_0}=1$ (up to numerical accuracy). The existence of the turnaround position is guaranteed by the symmetric configuration that we consider here. Below we show two such representative profiles for $\omega =0.5$ and $\omega=3$ respectively (corresponding to the blue and the red curve).

\begin{figure}[!ht]
\begin{center}
\includegraphics[angle=0,
width=0.65\textwidth]{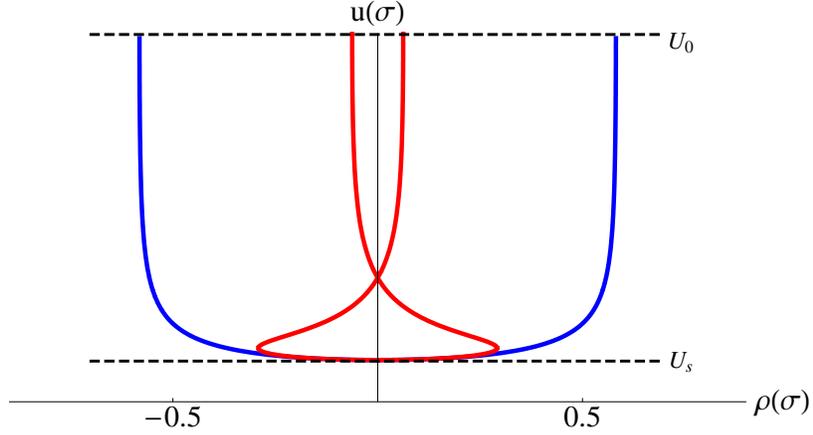}
\caption{\small Two representative spinning string profiles. We have set $R=1$.}
\label{fig: mdynT}
\end{center}
\end{figure}

Such profiles were discussed in ref.~\cite{Kruczenski:2003be} in a different but closely related model. In brief, we find that for a given value of $\omega$ there are classes of solutions distinguished by the number of times they cross zero (referred to as ``nodes" in ref.~\cite{Kruczenski:2003be}) along the horizontal axis. Here we will constrain ourselves to the $n=1$ case only.

We will now explore the behaviour of high spin mesons in presence of a background magnetic field. The key feature that we will observe for both zero and finite temperature phases is the existence of a shift in physical quantities such as energy and angular momentum in presence of a non-zero magnetic field. We begin with our numerical results for the zero temperature phase.

\subsubsection{Zero Temperature Phase}

\begin{figure}[!ht]
\begin{center}
\subfigure[] {\includegraphics[angle=0,
width=0.45\textwidth]{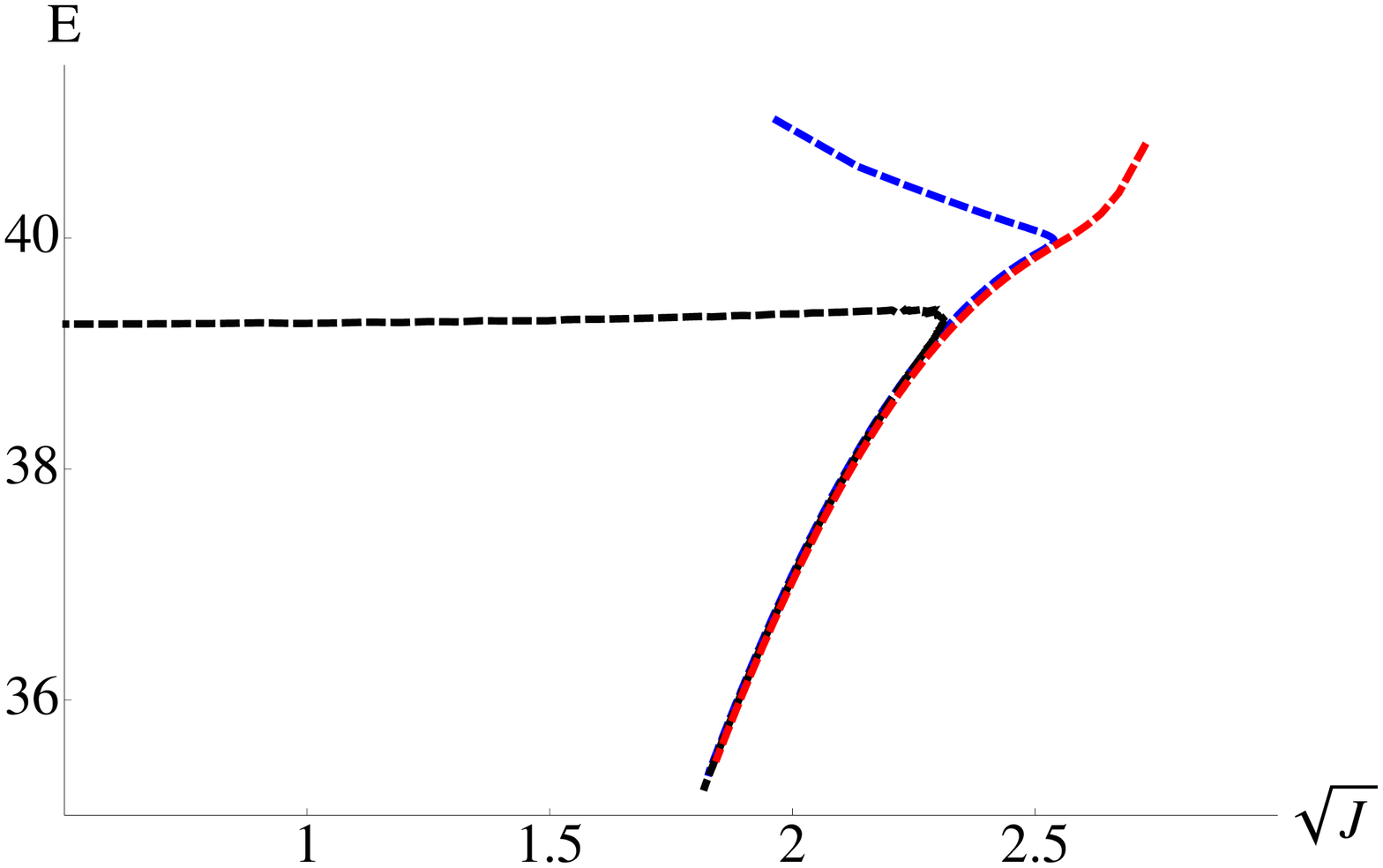} \label{fig: ej}}
\subfigure[] {\includegraphics[angle=0,
width=0.45\textwidth]{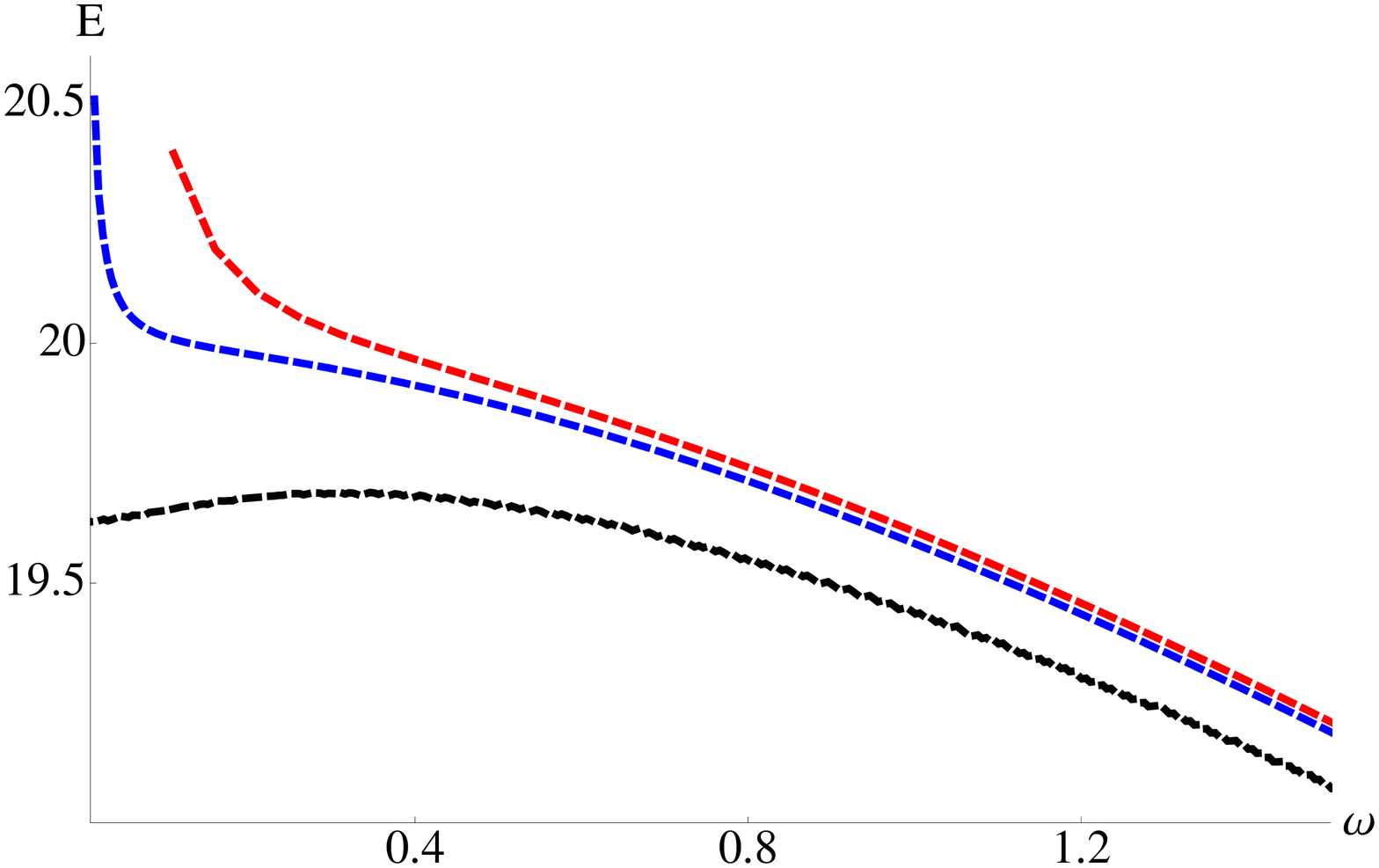} \label{fig: eo}}
\subfigure[] {\includegraphics[angle=0,
width=0.45\textwidth]{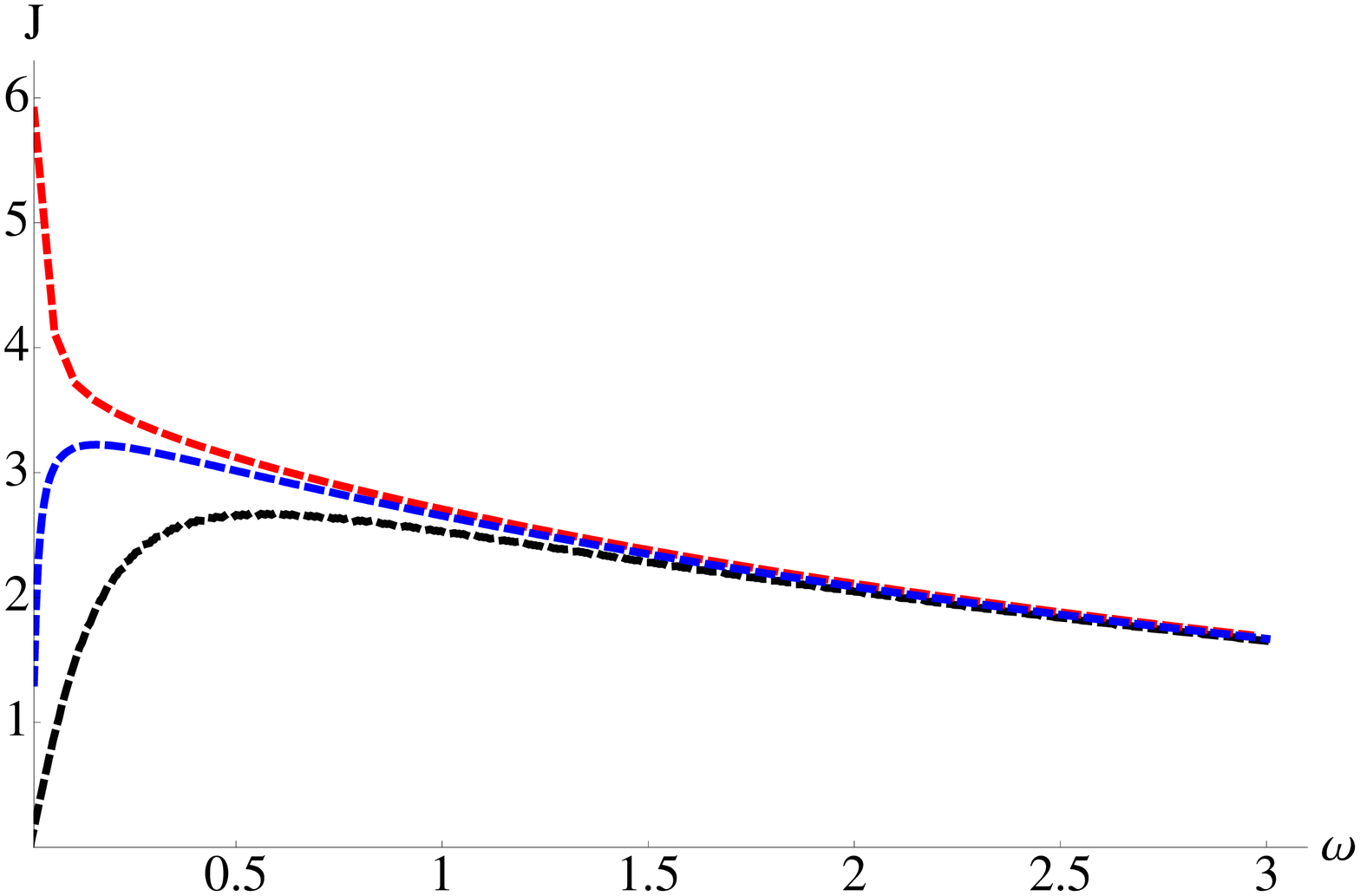} \label{fig: jo}}
\caption{\small The effect of an external magnetic field in the zero temperature phase. The black dashed lines represent $B=0$, blue dashed lines represent $B=-0.1$ and the red dashed lines represent $B=0.1$; $E$ and $J$ have been evaluated in units of $1/(2\pi\alpha')$. $B$ have been evaluated in units of $R$.}
\end{center}
\end{figure}\label{fig: can}

In figure \ref{fig: eo} we have shown the dependence of the meson energy $E$ as a function of the angular frequency $\omega$, in figure \ref{fig: jo} we have plotted the dependence of meson angular momentum  again as a function of $\omega$ and in figure \ref{fig: ej} we have shown the dependence of energy with angular momentum of the meson. For zero external magnetic field we recover the behaviour discussed in ref.~\cite{Peeters:2006iu}. We find that the external magnetic field shifts the energy and the angular momentum of the meson for generic values of the parameters of the system; however with increasing angular frequency these observables tend to become insensitive to the external field. In view of the approximate analytic results obtained in eqn. (\ref{eqt: splite}) it is straightforward to observe that an external field contributes only at order $B/\omega$ and hence has vanishingly small effects for large enough angular frequency. 

The existence of two energy branches for the same given value of $J$ seems to persist (as in vanishing external field) even in the presence of a magnetic field, although as the magnetic field is increased we observe a tendency that this multi-valuedness in energy starts disappearing. Therefore any possible unstable upper branch in figure \ref{fig: ej} can get promoted to a stable one by having sufficiently high angular momentum.  

Moreover we observe that there exists a maximum angular momentum beyond which the spinning meson dissociates. This dissociation is mediated by the acceleration of the spinning string. However the presence of magnetic field can again stabilize the mesons by raising the maxima in figure \ref{fig: jo}. Physically this simply corresponds to the fact that the magnetic field enhances the angular momentum of the meson and thus makes it stable.

\begin{figure}[!ht]
\begin{center}
\includegraphics[angle=0,
width=0.55\textwidth]{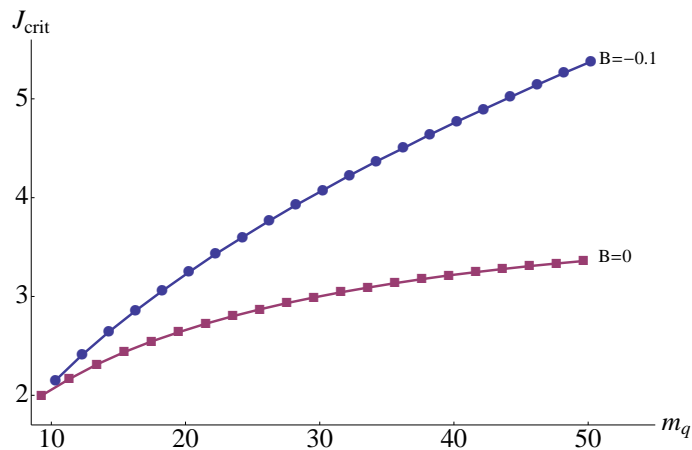}
\caption{\small Dependence of maximum spin with magnetic field for varying constituent quark mass. $J_{\rm crit}$ and $m_q$ have been calculated in units of $1/(2\pi\alpha')$. We observe that even for a certain negative value of $B$, i.e. when the magnetic field is aligned anti-parallel to its angular momentum, the meson can be more stable than its zero field counterpart.}
\label{fig: jcrit}
\end{center}
\end{figure}

To demonstrate this effect we observe the plot in figure \ref{fig: jcrit}. Clearly the $B=-0.1$ branch is higher than the $B=0$ branch; an even higher value of $B$ would inhibit any dissociation at all. Also the heavier the constituent quark mass is the more $J_{\rm crit}$ becomes making it less likely for the heavy mesons to dissociate.

\begin{figure}[!ht]
\begin{center}
\includegraphics[angle=0,
width=0.55\textwidth]{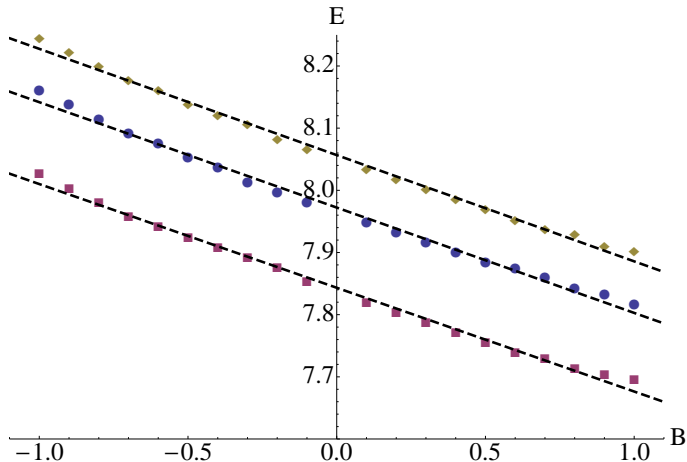}
\caption{\small Linear correction to energy for the equal charge case. From top to bottom the values of angular momenta are $J=0.937,$ $0.907,$ $0.862$ respectively. The dashed straight lines are the best fit lines. $E$ and $J$ are both presented in units of $1/(2\pi\alpha')$. We observe that all the three best fit lines have same (up to numerical accuracy) slope, which we expect from the very nature of Zeeman splitting.}
\label{fig: zee}
\end{center}
\end{figure}

A key feature of the high spin meson spectrum in presence of an external magnetic field is the linear Zeeman splitting. In figure \ref{fig: zee} we have shown such linear correction in energy for fixed values of the meson angular momentum.

\subsubsection{Finite Temperature Phase}

We now perform a similar numerical study of high spin mesons in this intermediate temperature phase.

\begin{figure}[!ht]
\begin{center}
\subfigure[] {\includegraphics[angle=0,
width=0.45\textwidth]{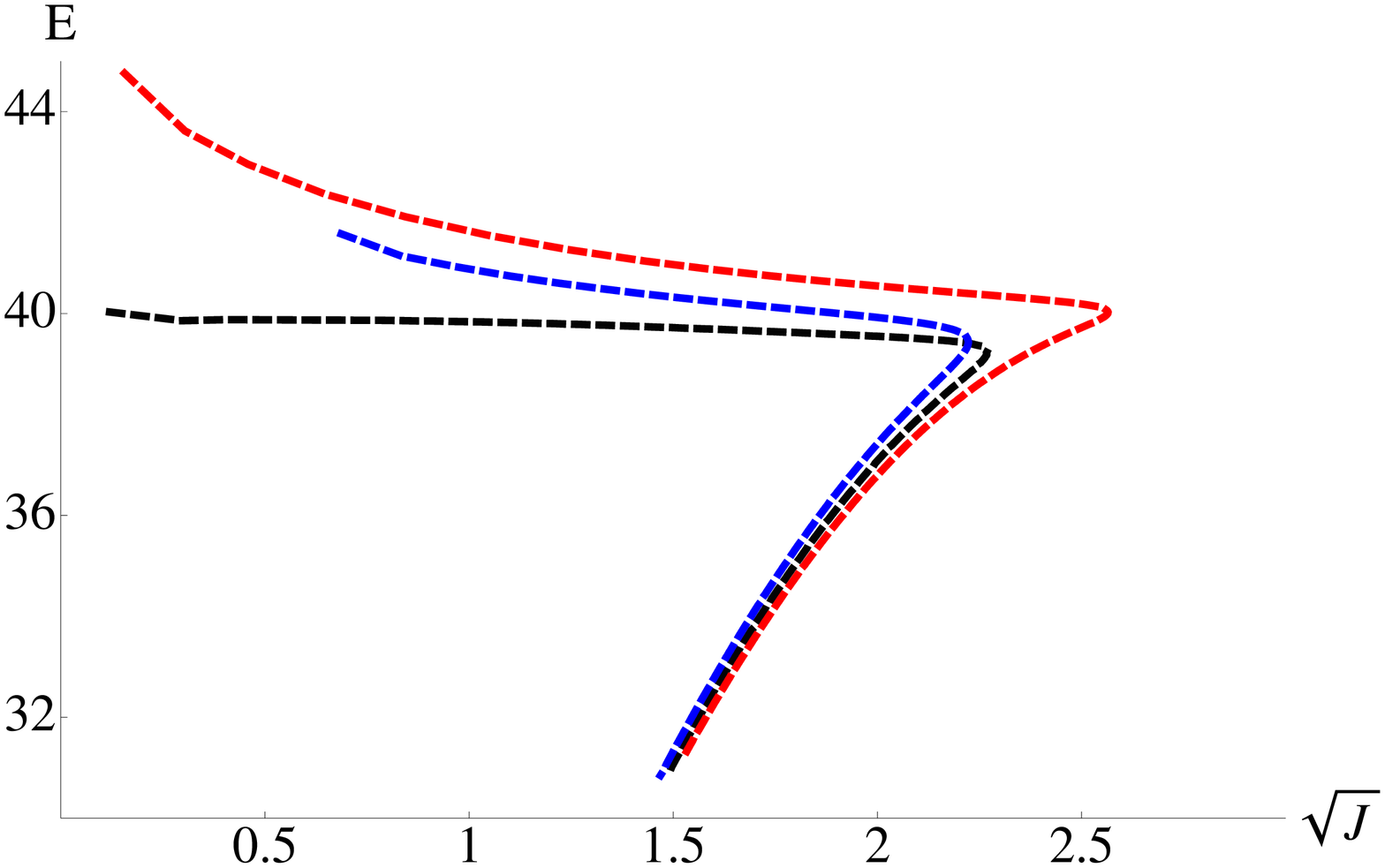} \label{fig: ejt}}
\subfigure[] {\includegraphics[angle=0,
width=0.45\textwidth]{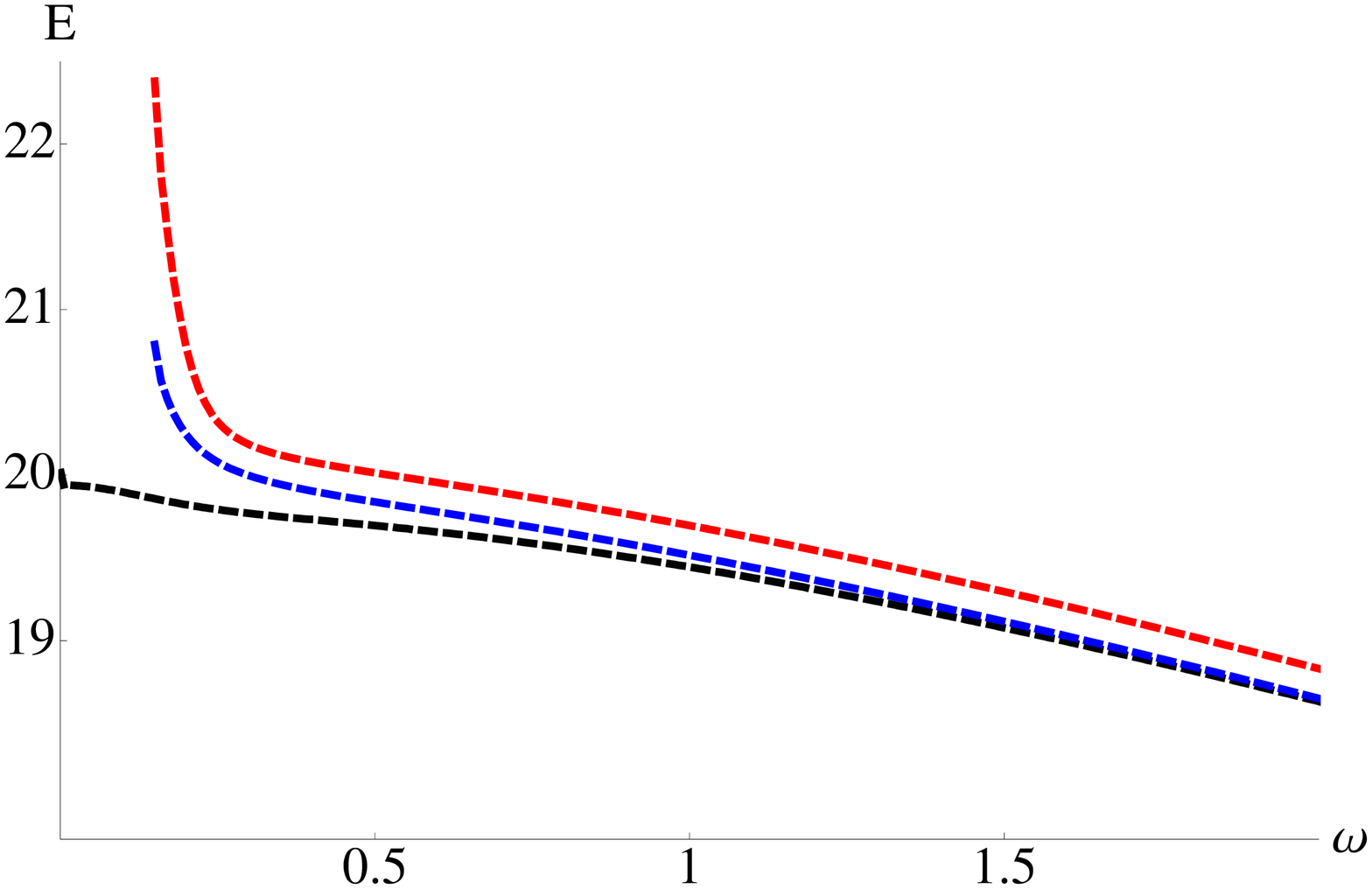} \label{fig: eot}}
\subfigure[] {\includegraphics[angle=0,
width=0.45\textwidth]{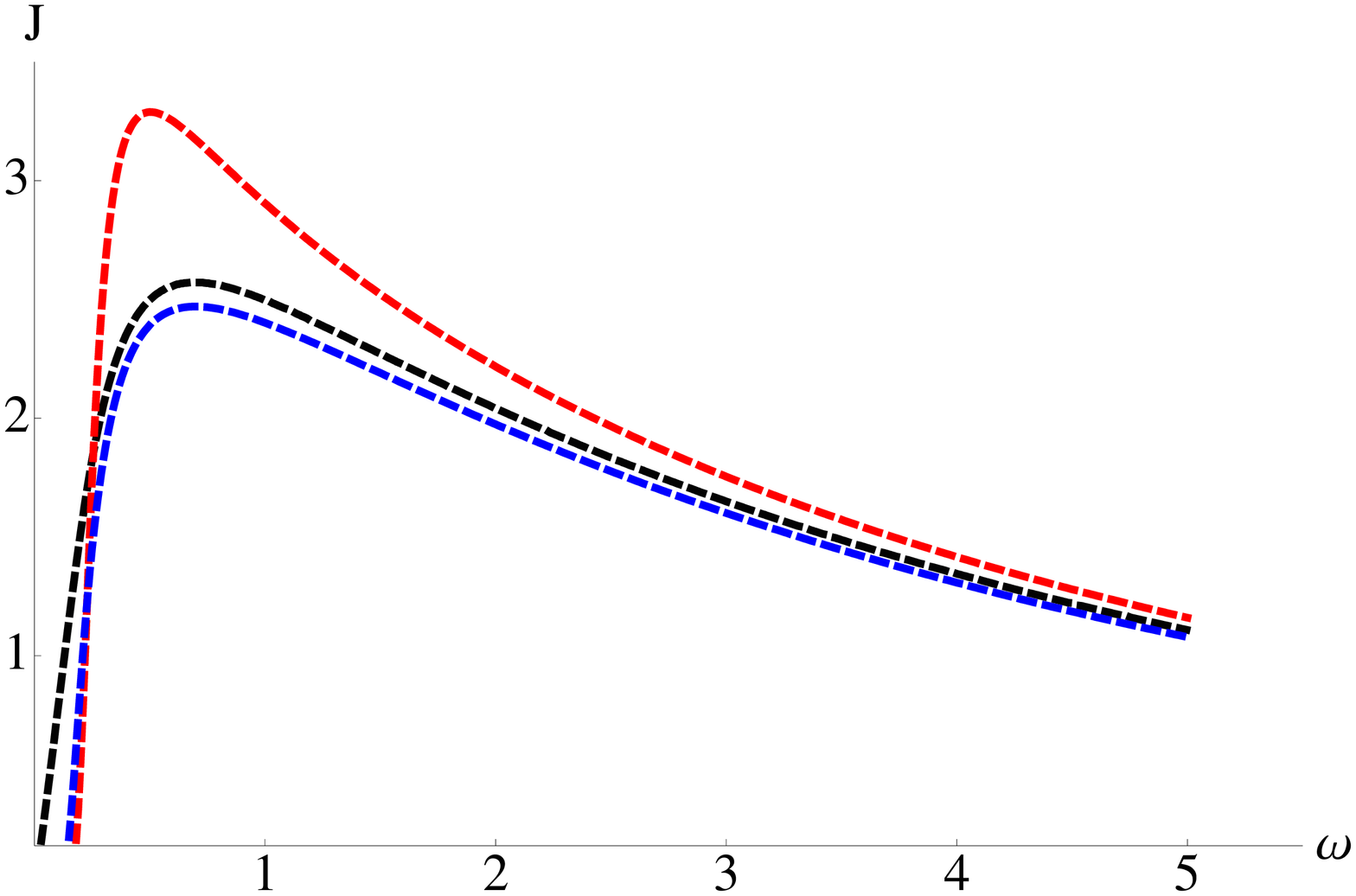} \label{fig: jot}}
\caption{\small The shifts in presence of an external magnetic field. $E$ and $J$ have been computed in units of $1/(2\pi\alpha')$. $B$ has been evaluated in units of $R$. The black curve corresponds to vanishing magnetic field, the blue curve corresponds to setting $B=-1$ and the red curve corresponds to setting $B=1$, where $B$ has been expressed in units of $R$.}
\end{center}
\end{figure}

In figure \ref{fig: eot} we have shown the dependence of meson energy as a function of its angular frequency $\omega$; in figure \ref{fig: jot} we have plotted the angular momentum of the meson as a function of $\omega$ and in figure \ref{fig: ejt} we have shown the dispersion relation between the meson energy and its angular momentum. The qualitative features of finite temperature physics are similar to that of the zero temperature physics. There is however one distinction as compared to the energy spectrum at zero temperature phase. The presence of magnetic field at finite temperature does not promote the possible unstable upper branch in figure \ref{fig: ejt} to a stable one within the range of values for the magnetic field that we have explored using our numerical approach (we expect however that for sufficiently high values of $B$ this unstable mode will be promoted to be a stable one). A possible thermally enhanced decay channel therefore remains open at finite temperature for a rather high value of the external field. The other familiar role of the magnetic field, we again find, is to introduce shifts in the physical quantities such as the energy and angular momentum of the meson. 

\begin{figure}[!ht]
\begin{center}
\subfigure[] {\includegraphics[angle=0,
width=0.45\textwidth]{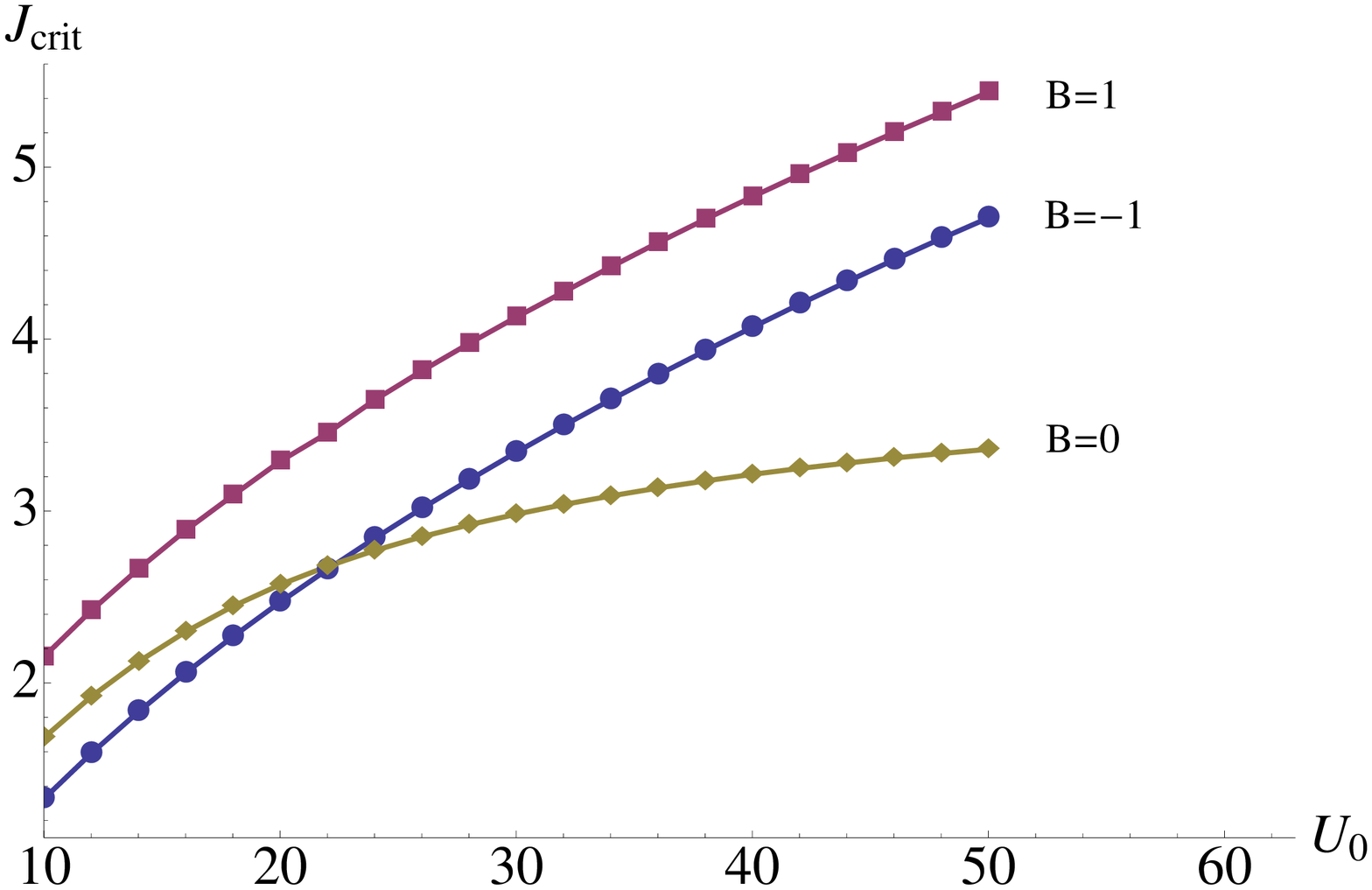} \label{fig: jcritt}}
\subfigure[] {\includegraphics[angle=0,
width=0.45\textwidth]{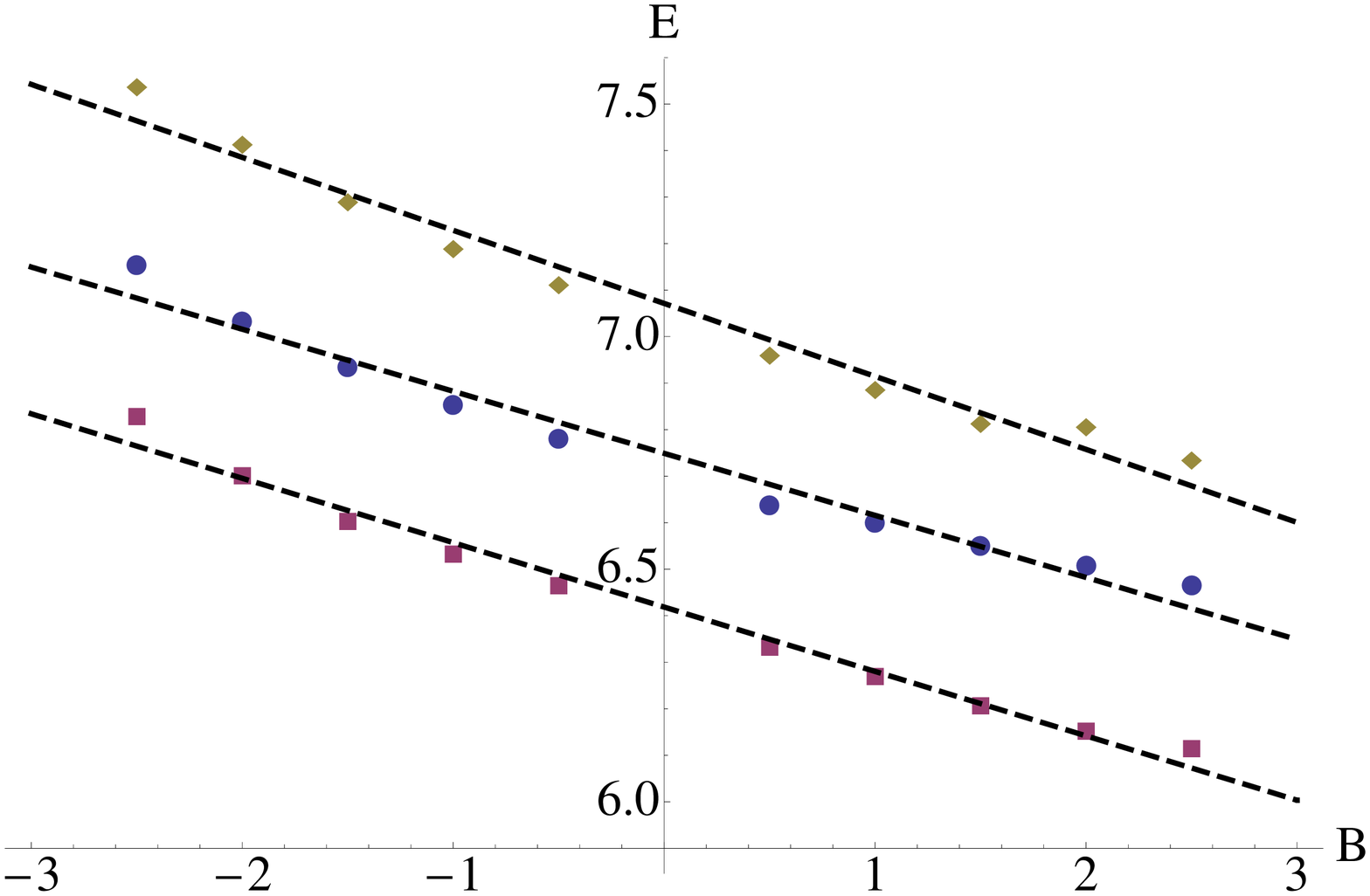} \label{fig: zeeT}}
\caption{\small Figure \ref{fig: jcritt} shows the dependence of maximum spin calculated in units of $1/(2\pi\alpha')$ with the applied magnetic field. We can observe that there exists a certain range of values for $U_0$, which in turn fixes the value of the constituent quark mass, where the angular momentum is lowered by the magnetic field. Figure \ref{fig: zeeT} shows the linear correction to meson energy for the equal charge case at finite temperature. From top to bottom the values of angular momenta are $J=0.630,$ $0.560$ and $0.490$ evaluated in units of $1/(2\pi\alpha')$ respectively. The dashed lines are the best fit lines.}
\end{center}
\end{figure}

The meson dissociation due to spin should now be enhanced due to thermal fluctuation. We have previously observed that the magnetic field adds extra angular momentum to the system and stabilizes the mesons. We study in figure \ref{fig: jcritt} the effect of two competing parameters on the critical angular momentum of the meson. Finally we observe the familiar Zeeman splitting in figure \ref{fig: zeeT}.

\section{Conclusions}

We have uncovered a sub--sector of the meson spectrum in the presence
of an external magnetic field, complementing the phase structure that
we obtained in ref.~\cite{Johnson:2008vna}. Mesons with small spin that
can be obtained by studying the quadratic fluctuations of the probe
brane classical configuration are harder to study in absolute
generality. In general the task is to solve a set of coupled
differential equations, which we have here analysed in the situations
where it is possible to decouple a subset of the modes. We have also
analysed the large spin meson spectra in presence of a magnetic field
to realize the well--known Zeeman effect. We also found that an external magnetic field enhances the stability of mesons and inhibits the dissociation. In appendix D we have
presented a model calculation analyzing spinning strings in Rindler
space to capture key features of the large spin meson dissociation.

There are several directions to pursue in future work. It is of
interest to obtain the low spin meson spectra in the low temperature
phase ({\it i.e.,} when chiral symmetry is broken) and their
dispersion relations. In the present work, we have focussed on the
effect of a magnetic field. The analysis of the meson spectra and
quasinormal modes in the presence of an electric field could be
pursued in a similar spirit. The effect on the spectra of the presence
of a chemical potential would also be of great interest. These
external fields can all be realized as the non--normalizable modes of
anti--symmetric fields on the world volume of the probe brane, and so
we would expect that in all these cases in the high temperature
chirally symmetric phase, the scalar and vector fluctuations always
remain decoupled, as we have seen here.

\section*{Acknowledgments}
AK would like to thank the Tata Institute of Fundamental Research for hospitality during the Monsoon Workshop on String Theory where a part of this work was carried out. AK would also like to thank the International Center for Theoretical Sciences, India for hospitality and support. This work was supported by the US Department of Energy.

\renewcommand{\theequation}{A-\arabic{equation}}
\setcounter{equation}{0}  
\section*{Appendix A}  

Here we provide the details of the computation for quadratic action for fluctuation of the probe brane. To keep the story general we consider the scalar and vector fluctuation simultaneously since in presence of an external magnetic field they are likely to couple. Parametrizing the worldvolume of the $D8/\overline{D8}$-brane by the function $\tau_0(u)$, we consider the following fluctuation
\begin{eqnarray}
\tau(u)=\tau_0(u)+(2\pi\alpha')\chi(\xi^a)\ ,
\end{eqnarray}
where $\tau_0(u)$ is the classical embedding of the probe brane and $\{\xi^a\}$ refer to its worldvolume coordinates. 

To proceed let us recall that the DBI lagrangian is given by 
\begin{eqnarray}
\mathcal{L}_{\rm DBI}&=&e^{-\phi}\sqrt{-{\rm det}\left({\rm P}[G_{\mu\nu}+B_{\mu\nu}]+(2\pi\alpha' F_{ab})\right)}\nonumber\\
                                &=&e^{-\phi}\sqrt{-{\rm det}\left(E_{ab}^{(0)}+(2\pi\alpha')E_{ab}^{(1)}+(2\pi\alpha')^2E_{ab}^{(2)}\right)}\nonumber\\
                                &=&e^{-\phi}\sqrt{-{\rm det}\left(E_{ab}^{(0)}\right)}\left(1+\frac{1}{2}{\rm Tr }\mathcal{M}-\frac{1}{4}{\rm Tr} \mathcal{M}^2+\frac{1}{8}({\rm Tr} \mathcal{M})^2\right)+\mathcal{O}\left(\mathcal{M}^3\right)\ ,
\end{eqnarray}
where 
\begin{eqnarray}
&& \mathcal{M}_a^c=(2\pi\alpha')\left(E_{cb}^{(0)}\right)^{-1}E_{ab}^{(1)}+(2\pi\alpha')^2\left(E_{cb}^{(0)}\right)^{-1}E_{ab}^{(2)}\ ,\nonumber\\
&& E_{ab}^{(0)}=G_{\mu\nu}\left(\partial_a X^\mu\right)\left(\partial_b X^\nu\right)+B_{ab}\ ,\quad
 E_{ab}^{(1)}=F_{ab}+G_{\tau\tau}\tau_0'(u)\left[\delta_b^u\left(\partial_a\chi\right)+\delta_a^u\left(\partial_b\chi\right)\right]\ ,\nonumber\\
&& E_{ab}^{(2)}=G_{\tau\tau}\left(\partial_a\chi\right)\left(\partial_b\chi\right)\ .
\end{eqnarray}
Here $G_{\mu\nu}$ and $\{X^\mu\}$ refer to the bulk spacetime. For organizational convenience let us also note that
\begin{eqnarray}\label{eqt: comp}
&& \left(E_{ab}^{(0)}\right)^{-1}=\mathcal{S}^{ab}+\mathcal{A}^{ab}\ ,\quad
 \mathcal{S}^{ab}={\rm diag}\left\{-G^{tt},G^{xx},\frac{G_{xx}}{G_{xx}^2+B^2},\frac{G_{xx}}{G_{xx}^2+B^2},g_{uu}^{-1}\right\}\times \left |\left |\Omega_4\right |\right |\ ,\nonumber\\
&& \mathcal{A}^{ab}=\frac{B}{G_{xx}^2+B^2}\left(\delta_3^a\delta_2^b-\delta_2^a\delta_3^b\right)\ , \quad g_{uu}=G_{uu}+\tau_0'(u)^2G_{\tau\tau}\ , \nonumber\\
&& \left |\left |\Omega_4\right |\right |={\rm diag}\left\{\left(u/R\right)^{-3/2}u^2, \left(u/R\right)^{-3/2}u^2,\left(u/R\right)^{-3/2}u^2,\left(u/R\right)^{-3/2}u^2\right\}\ ,\nonumber\\
&& \sqrt{-{\rm det}\left(E_{ab}^{(0)}\right)}=\sqrt{G_{tt}G_{xx}g_{uu}\left(G_{xx}^2+B^2\right)\left({\rm det} \left |\left |\Omega_4\right |\right |\right)}\ ,
\end{eqnarray}
where $\mathcal{S}$ and $\mathcal{A}$ are the symmetric and anti-symmetric part respectively and $\left |\left |\Omega_4\right |\right |$ denotes the diagonal metric for the $S_4$. 

For the low temperature background we have
\begin{eqnarray}
G_{\tau\tau}=\left(u/R\right)^{3/2}f(u)\ ,\quad G_{xx}=G_{tt}=\left(u/R\right)^{3/2}\ ,\quad G_{uu}=\left(\frac{u}{R}\right)^{-3/2}\frac{1}{f(u)}\ , f(u)=1-\left(\frac{u_{KK}}{u}\right)^3\ , \nonumber
\end{eqnarray}
and for the high temperature background we have
\begin{eqnarray}
G_{tt}=\left(u/R\right)^{3/2}f(u)\ ,\quad G_{xx}=G_{\tau\tau}=\left(u/R\right)^{3/2}\ ,\quad G_{uu}=\left(\frac{u}{R}\right)^{-3/2}\frac{1}{f(u)}\ , f(u)=1-\left(\frac{u_{T}}{u}\right)^3\ . \nonumber
\end{eqnarray}
In what follows we will set $R=1$.

Now we can determine the effective lagrangian corresponding to small fluctuations. The contribution at order $(2\pi\alpha')^2$ is summarised below
\begin{eqnarray}
&& S_{\rm DBI} = \int d^9\xi \mathcal{L}_{\rm total}\ , \quad \mathcal{L}_{\rm total}=-\mu_8\left[\mathcal{L}_{\chi}+\mathcal{L}_{F}+\mathcal{L}_{\chi F}\right]\ ,\nonumber\\
&& \mathcal{L}_{\chi}=\frac{1}{2}G_{\tau\tau}\left(\mathcal{S}^{uu}G_{\tau\tau}\tau_0'(u)^2-1\right)e^{-\phi}\sqrt{-{\rm det}\left(E_{ab}^{(0)}\right)}\mathcal{S}^{ab}\left(\partial_a\chi\right)\left(\partial_b\chi\right)\ ,\nonumber\\
&& \mathcal{L}_{F}=\frac{1}{4}e^{-\phi}\sqrt{-{\rm det}\left(E_{ab}^{(0)}\right)}\mathcal{S}^{aa'}\mathcal{S}^{bb'}F_{ab}F_{a'b'}\ ,\nonumber\\
&& \mathcal{L}_{\chi F}=\frac{1}{2}\partial_u\left[e^{-\phi}\tau_0'(u)G_{\tau\tau}\mathcal{S}^{uu}\sqrt{-{\rm det}\left(E_{ab}^{(0)}\right)} \mathcal{A}^{ab}\right]\chi F_{ab}\ .
\end{eqnarray}
The term $\mathcal{L}_{\chi F}$ is the interaction term that couples the scalar and the vector modes. Clearly if $\tau_0'(u)=0$ then the coupling vanishes. Therefore for the high temperature phase (i.e., when $\tau_0(u)={\rm const.}$) the scalar and vector meson modes are always decoupled. 

Now we determine the contribution coming from the Wess-Zumino term. To this end let us first recall that the background has a $C_5$ potential and a $C_3$ potential dual to $C_5$. The $C_5$ potential is given by
\begin{eqnarray}
C_5=g_s^{-1}\left(\frac{u}{R}\right)^3dx^0\wedge dx^1\wedge dx^2\wedge dx^3\wedge d\tau\ .
\end{eqnarray}
On the worldvolume of the probe brane this $C_5$ induces a $5$-form given by 
\begin{eqnarray}
P[C_5]=g_s^{-1}\left(\frac{u}{R}\right)^3\left(\tau_0'(u)+2\pi\alpha'\chi\right)dx^0\wedge dx^1\wedge dx^2\wedge dx^3\wedge du \ .
\end{eqnarray}
Therefore we can immediately observe that for the high temperature phase this cannot contribute at the leading order. 

Let us now recall that the Wess-Zumino term is given by
\begin{eqnarray}
S_{\rm WZ}=\mu_8\int_{\mathcal{M}_9}\sum_p C_{(p+1)}\wedge {\rm exp}\left(2\pi\alpha' F+B\right)\ .
\end{eqnarray}
It is straightforward to see that ${\rm P}[C_{(5)}]\wedge B_{(2)}$ vanishes identically. Therefore the total contribution coming from the Wess-Zumino term at order $(2\pi\alpha')^2$ is given by
\begin{eqnarray}\label{eqt: lwz}
S_{\rm WZ}&=&\frac{1}{2}\mu_8\int_{\mathcal{M}_9}\left({\rm P}[C_5]\wedge F_2\wedge F_2-\frac{1}{3} F_4\wedge A\wedge F_2\wedge B_2\right)\ , \quad {\rm with}\nonumber\\
F_4&=&dC_3\, \quad F_2=dA\ ,
\end{eqnarray}
We have performed an integration by parts to obtain the second term above. In what follows we will focus on the special case of $\tau_0'(u)=0$ in which case the first term in eqn. (\ref{eqt: lwz}) does not contribute at the leading order.

The equations of motion for the gauge fields then decouple from the scalar fluctuation and is given by (using $R=1$)
\begin{eqnarray}\label{eqt: aeom}
\partial_a\left[e^{-\phi}\sqrt{-{\rm det}\left(E_{ab}^{(0)}\right)}\mathcal{S}^{aa'}\mathcal{S}^{bb'}F_{a'b'}\right]+ 4 g_s^{-1} B\epsilon^{ba'b'23}\partial_{b'} A_{a'}=0\ .
\end{eqnarray}
where now $a, a', b, b'\in \{\mathbb{R}^{1,3}\}\bigcup \{u\}$, on the probe brane worldvolume. It is now possible to consistently set all the $A_\alpha=0$ where $\alpha\in S^4$. This choice imposes a constraint of the following form
\begin{eqnarray}\label{eqt: constr2}
\mathcal{S}^{\mu\nu}\partial_\alpha\partial_{\mu}A_{\nu}=0\ .
\end{eqnarray}
This constraint can be trivially satisfied by looking at the gauge field fluctuations independent of the spherical directions (in other words focussing only on the $SO(5)$ singlet states).

The most general ansatz for the gauge field fluctuations consistent with the previous constraint is given by
\begin{eqnarray}
A_t&=&A_t(u)e^{-i\omega t+ik x^1+ik_2x^2+ik_3x^3}\ , \nonumber\\
A_i&=&A_i(u)e^{-i\omega t+ik x^1+ik_2x^2+ik_3x^3}\ , \nonumber\\
A_u&=&A_u(u)e^{-i\omega t+ik x^1+ik_2x^2+ik_3x^3}\ .
\end{eqnarray}

To fix the residual gauge, we further impose $A_u=0$. This gauge choice gives the following constraint
\begin{eqnarray}\label{eqt: constrA}
e^{-\phi}\sqrt{-{\rm det}\left(E_{ab}^{(0)}\right)}\mathcal{S}^{uu}\left[\mathcal{S}^{tt}(\omega A_t')-\mathcal{S}^{11}(kA_1')-\mathcal{S}^{22}(k_2A_2')-\mathcal{S}^{33}(k_3A_3')\right]\nonumber\\
+4 g_s^{-1} B (\omega A_1+kA_t)=0\ .
\end{eqnarray}

With this ansatz the equations of motion for the gauge fields are given by
\begin{eqnarray}\label{eqt: eomA}
&& \partial_u\left[e^{-\phi}\sqrt{-{\rm det}\left(E_{ab}^{(0)}\right)}\mathcal{S}^{tt}\mathcal{S}^{uu}A_t'\right]+4 g_s^{-1} B A_1' \nonumber\\
&& -e^{-\phi}\sqrt{-{\rm det}\left(E_{ab}^{(0)}\right)} \mathcal{S}^{tt}\left[k\mathcal{S}^{11}(\omega A_1+kA_t)+k_2\mathcal{S}^{22}(\omega A_2+k_2A_t)+k_3\mathcal{S}^{33}(\omega A_3+k_3A_t)\right]=0\ , \nonumber\\
&& \partial_u\left[e^{-\phi}\sqrt{-{\rm det}\left(E_{ab}^{(0)}\right)}\mathcal{S}^{11}\mathcal{S}^{uu}A_1'\right]-4 g_s^{-1} B A_t' \nonumber\\
&& -e^{-\phi}\sqrt{-{\rm det}\left(E_{ab}^{(0)}\right)} \mathcal{S}^{11}\left[k_2\mathcal{S}^{22}(k_2A_1-kA_2)+k_3\mathcal{S}^{33}(k_3A_1-kA_3)+\omega\mathcal{S}^{tt}(\omega A_1+kA_t)\right]=0\ , \nonumber\\
&& \partial_u\left[e^{-\phi}\sqrt{-{\rm det}\left(E_{ab}^{(0)}\right)}\mathcal{S}^{22}\mathcal{S}^{uu}A_2'\right]\nonumber\\
&& -e^{-\phi}\sqrt{-{\rm det}\left(E_{ab}^{(0)}\right)} \mathcal{S}^{22}\left[k\mathcal{S}^{11}(kA_2-k_2A_1)+k_3\mathcal{S}^{33}(k_3A_2-k_2A_3)+\omega\mathcal{S}^{tt}(\omega A_2+k_2A_t)\right]=0\ , \nonumber\\
&& \partial_u\left[e^{-\phi}\sqrt{-{\rm det}\left(E_{ab}^{(0)}\right)}\mathcal{S}^{33}\mathcal{S}^{uu}A_3'\right]\nonumber\\
&& -e^{-\phi}\sqrt{-{\rm det}\left(E_{ab}^{(0)}\right)} \mathcal{S}^{33}\left[k\mathcal{S}^{11}(kA_3-k_3A_1)+k_2\mathcal{S}^{22}(k_2A_3-k_3A_2)+\omega\mathcal{S}^{tt}(\omega A_3+k_3A_t)\right]=0\ . \nonumber\\
\end{eqnarray}

In general, the gauge field fluctuations can have momentum along any of the spatial directions. Due to the presence of the magnetic field there are two particularly interesting cases to consider. The fluctuation field can either oscillate in the direction parallel to the magnetic field, or in the direction perpendicular to the magnetic field. We discuss these cases below.\\

\textbf{Case 1: Oscillation parallel to the magnetic field} \\

Let us consider first restricting ourselves to $k_2=0=k_3$. Equations in (\ref{eqt: eomA}) then reduce to
\begin{eqnarray}\label{eqt: eomA1}
&& \partial_u\left[e^{-\phi}\sqrt{-{\rm det}\left(E_{ab}^{(0)}\right)}\mathcal{S}^{tt}\mathcal{S}^{uu}A_t'\right]+4 g_s^{-1} B A_1' -e^{-\phi}\sqrt{-{\rm det}\left(E_{ab}^{(0)}\right)} \mathcal{S}^{tt}k\mathcal{S}^{11}(\omega A_1+kA_t)=0\ , \nonumber\\
&& \partial_u\left[e^{-\phi}\sqrt{-{\rm det}\left(E_{ab}^{(0)}\right)}\mathcal{S}^{11}\mathcal{S}^{uu}A_1'\right]-4 g_s^{-1} B A_t' -e^{-\phi}\sqrt{-{\rm det}\left(E_{ab}^{(0)}\right)} \mathcal{S}^{11}\omega\mathcal{S}^{tt}(\omega A_1+kA_t)=0\ , \nonumber\\
&& \partial_u\left[e^{-\phi}\sqrt{-{\rm det}\left(E_{ab}^{(0)}\right)}\mathcal{S}^{22}\mathcal{S}^{uu}A_2'\right] -e^{-\phi}\sqrt{-{\rm det}\left(E_{ab}^{(0)}\right)} \mathcal{S}^{22}\left(k^2\mathcal{S}^{11}+\omega^2\mathcal{S}^{tt}\right)A_2=0\ , \nonumber\\
&& \partial_u\left[e^{-\phi}\sqrt{-{\rm det}\left(E_{ab}^{(0)}\right)}\mathcal{S}^{33}\mathcal{S}^{uu}A_3'\right] -e^{-\phi}\sqrt{-{\rm det}\left(E_{ab}^{(0)}\right)} \mathcal{S}^{33}\left(k^2\mathcal{S}^{11}+\omega^2\mathcal{S}^{tt}\right)A_3=0\ ,
\end{eqnarray}
along with the following constraint
\begin{equation}\label{eqt: constrpara}
e^{-\phi}\sqrt{-{\rm det}\left(E_{ab}^{(0)}\right)}\mathcal{S}^{uu}\left[\mathcal{S}^{tt}(\omega A_t')-\mathcal{S}^{11}(kA_1')\right]+4 g_s^{-1} B (\omega A_1+kA_t)=0\ .
\end{equation}

In this case, the longitudinal modes ($A_t, A_1$ components oscillating along $\{t,x^1\}$ plane) and the transverse modes ($A_2, A_3$ components oscillating along $\{t,x^1\}$ plane) clearly decouple. It is useful to note that $A_2$, $A_3$ transform as vectors and $\{A_t, A_1\}$ transforms as a scalar under the $SO(2)$ rotation group in the plane perpendicular to the magnetic field.\\

\textbf{Case 2: Oscillation perpendicular to the magnetic field} \\

Now we restrict ourselves to $k=0=k_2$. In this case the equations of motion in (\ref{eqt: eomA}) become
\begin{eqnarray}
&& \partial_u\left[e^{-\phi}\sqrt{-{\rm det}\left(E_{ab}^{(0)}\right)}\mathcal{S}^{tt}\mathcal{S}^{uu}A_t'\right]+4 g_s^{-1} B A_1' -e^{-\phi}\sqrt{-{\rm det}\left(E_{ab}^{(0)}\right)} \mathcal{S}^{tt}k_3\mathcal{S}^{33}(\omega A_3+k_3A_t)=0\ , \nonumber\\
&& \partial_u\left[e^{-\phi}\sqrt{-{\rm det}\left(E_{ab}^{(0)}\right)}\mathcal{S}^{11}\mathcal{S}^{uu}A_1'\right]-4 g_s^{-1} B A_t' -e^{-\phi}\sqrt{-{\rm det}\left(E_{ab}^{(0)}\right)} \mathcal{S}^{11}\left(k_3^2\mathcal{S}^{33}+\omega^2\mathcal{S}^{tt}\right)A_1=0\ , \nonumber\\
&& \partial_u\left[e^{-\phi}\sqrt{-{\rm det}\left(E_{ab}^{(0)}\right)}\mathcal{S}^{33}\mathcal{S}^{uu}A_3'\right] -e^{-\phi}\sqrt{-{\rm det}\left(E_{ab}^{(0)}\right)} \mathcal{S}^{33}\omega\mathcal{S}^{tt}\left(\omega A_3+k_3A_t\right)=0\ , \nonumber\\
&& \partial_u\left[e^{-\phi}\sqrt{-{\rm det}\left(E_{ab}^{(0)}\right)}\mathcal{S}^{22}\mathcal{S}^{uu}A_2'\right] -e^{-\phi}\sqrt{-{\rm det}\left(E_{ab}^{(0)}\right)} \mathcal{S}^{22}\left(k_3^2\mathcal{S}^{33}+\omega^2\mathcal{S}^{tt}\right)A_2=0\ ,
\end{eqnarray}
along with the constraint
\begin{equation}
e^{-\phi}\sqrt{-{\rm det}\left(E_{ab}^{(0)}\right)}\mathcal{S}^{uu}\left[\mathcal{S}^{tt}(\omega A_t')-\mathcal{S}^{33}(k_3A_3')\right]\nonumber\\
+4 g_s^{-1} B (\omega A_1+kA_t)=0\ .
\end{equation}

In this case only the $A_2$-transverse mode decouple from the rest of the gauge field fluctuations. The longitudinal modes ($A_t, A_3$ components oscillating along $\{t,x^3\}$ plane) and the transverse mode ($A_1$ component oscillating along $\{t,x^3\}$ plane) are in general coupled.

\renewcommand{\theequation}{B-\arabic{equation}}
\setcounter{equation}{0}  
\section*{Appendix B}  

Here we analyze the gauge field fluctuations when there is no magnetic field. We provide an analytical derivation for the dispersion relation of the lowest hydrodynamic mode of the longitudinal oscillation, which has been obtained by other methods in refs. \cite{Myers:2007we, Evans:2008tv}.

The longitudinal mode is defined to be $\mathcal{E}(u)=\omega A_1+ kA_t$. Using this definition, the constraint equation in (\ref{eqt: constrpara}) and the equations of motion in (\ref{eqt: eomA1}) (after setting $B=0$) we obtain the equation of motion for the longitudinal mode to be
\begin{eqnarray}
&& \frac{\mathcal{E}(u) \left(\omega ^2-k^2 f(u)\right) R^3}{u^3 f(u)^2}+\frac{\mathcal{E}'(u) \left(5 f(u) \omega ^2+2 u f'(u) \omega^2-5 k^2 f(u)^2\right)}{2 u f(u) \left(\omega ^2-k^2 f(u)\right)}+\mathcal{E}''(u)=0\ ,\nonumber\\
&& f(u)=1-\left(\frac{u_T}{u}\right)^3\ .
\end{eqnarray}

With the following variable change
\begin{eqnarray}
x=\frac{u}{u_T}\ , 
\end{eqnarray}
the equation of motion for the longitudinal mode becomes
\begin{eqnarray}\label{eqt: h0eom}
\frac{d^2\mathcal{E}}{dx^2}+\frac{5\left(1-x^{-3}\right)\left(\omega^2-k^2\left(1-x^{-3}\right)\right)+6\omega^2x^{-3}}{2x\left(1-x^{-3}\right)\left(\omega^2-k^2\left(1-x^{-3}\right)\right)}\frac{d\mathcal{E}}{dx}+\frac{R^3\left(\omega^2-k^2\left(1-x^{-3}\right)\right)}{u_T x^3\left(1-x^{-3}\right)^2}\mathcal{E}(x)=0\ .\nonumber\\
\end{eqnarray}

The near-horizon limit $u\to u_T$ is now achieved by taking the $x\to 1$ limit. In this limit, the equation of motion takes the form
\begin{eqnarray}
\frac{d^2\mathcal{E}}{dx^2}+\frac{1}{x-1}\frac{d\mathcal{E}}{dx}+\frac{R^3\omega^2}{9u_T(x-1)^2}\mathcal{E}(x)=0\ .
\end{eqnarray}
The general solution of this equation is given by a linear combination of the incoming and the outgoing modes
\begin{eqnarray}
\mathcal{E}(x)=C_1 \cos (\tilde{\omega} \log (x-1))+C_2 \sin (\tilde{\omega}\log (x-1))\ , \quad \tilde{\omega}=\sqrt{\frac{R^3\omega^2}{9 u_T}}=\frac{\omega}{4\pi T}\ .
\end{eqnarray}
The incoming boundary condition singles out the solution with $C_1=-C_2$. Near the horizon $\mathcal{E}(x)$ is therefore obtained to be
\begin{eqnarray}\label{eqt: Ehori}
\mathcal{E}(x)=C_L {\rm Exp}\left(-i\tilde{\omega}\log(x-1)\right)\ ,
\end{eqnarray}
where $C_L$ is an yet undetermined constant.

Let us also define $\tilde{k}=k/(4\pi T)$. Now, for small enough $\omega$ and $k$ (such that $\tilde{\omega}\ll 1$ and $\tilde{k}\ll 1$), we can ignore the last term in equation (\ref{eqt: h0eom}). It turns out that in this limit the equation of motion for the longitudinal mode is also exactly solvable.
\begin{eqnarray}\label{eqt: eomE}
\mathcal{E}(x)=M_2+\frac{1}{3} M_1 \left(-\frac{2 \tilde{k}^2}{x^{3/2}}+\tilde{\omega} ^2 \log \left[\frac{x^{3/2}+1}{x^{3/2}-1}\right]\right)\ ,
\end{eqnarray}
where $M_1$ and $M_2$ are constants of integration. Near the boundary, i.e., $x\to\infty$, this reduces to
\begin{eqnarray}
\mathcal{E}(x)=M_2-\frac{2}{3}M_1\left(\tilde{k}^2-\tilde{\omega}^2\right)\left(\frac{1}{x^{3/2}}\right)\ .
\end{eqnarray}
Therefore normalizability of $\mathcal{E}(x)$ forces us to impose $M_2=0$. The quasi-normal modes are therefore obtained to be the solution of this constraint.

On the other hand, in the vicinity of the horizon the solution in (\ref{eqt: eomE}) takes the following form
\begin{eqnarray}\label{eqt: expan1}
\mathcal{E}(x)&=&\frac{1}{3} \left(-2 M_1 \tilde{k}^2+3 M_2+M_1 \tilde{\omega} ^2 \log \left(\frac{4}{3}\right)\right)-\frac{1}{3}M_1 \tilde{\omega} ^2 \log (x-1)+\nonumber\\
&& \frac{1}{6} M_1 \left(6 \tilde{k}^2+\tilde{\omega} ^2\right) (x-1)-\frac{1}{144} M_1 \left(180 \tilde{k}^2+\tilde{\omega} ^2\right) (x-1)^2+ \ldots
\end{eqnarray}

Now, for sufficiently small values of $\tilde{\omega}$, we can expand the solution obtained in (\ref{eqt: Ehori}) to get
\begin{eqnarray}\label{eqt: expan2}
\mathcal{E}(x)=C_L-i C_L\tilde{\omega}\log(x-1)-\frac{1}{2}C_L\tilde{\omega}^2\left(\log(x-1)\right)^2\ldots
\end{eqnarray}
Comparing equation (\ref{eqt: expan1}) and (\ref{eqt: expan2}) we get
\begin{eqnarray}
&& \frac{1}{3} \left(-2 M_1 \tilde{k}^2+3 M_2+M_1 \tilde{\omega} ^2 \log \left(\frac{4}{3}\right)\right)=C_L\ , \nonumber\\
&& \frac{1}{3}M_1 \tilde{\omega} ^2=i C_L\tilde{\omega}\ .
\end{eqnarray}
Setting $M_2=0$, the solution of this equation is given by
\begin{eqnarray}\label{eqt: dis0}
M_1=\frac{3iC_L}{\tilde{\omega}}\ , \quad \tilde{\omega}=-2i \tilde{k}^2+\dots 
\end{eqnarray}
Restating this result in dimensionful parameters, we get the lowest hydrodynamic quasinormal frequency
\begin{eqnarray}
\omega=-i D_R k^2\ , \quad D_R=\frac{1}{2\pi T}\ ,
\end{eqnarray}
where $D_R$ is the R-charge diffusion constant. This value for the diffusion constant was obtained numerically by studying the spectral functions of holographic flavours in ref.~\cite{Myers:2007we} and was also confirmed in ref.~\cite{Evans:2008tv} by numerically obtaining the lowest quasinormal mode of the longitudinal oscillation.

In the vanishing magnetic field limit, the transverse modes obey the following equations (obtained from eqn. (\ref{eqt: eomA1}))
\begin{eqnarray}\label{eqt: a23t}
&& \frac{A_2(u) \left(\omega ^2-k^2 f(u)\right) R^3}{u^3 f(u)^2}+\frac{A_2'(u) \left(5 f(u)+2u f'(u)\right)}{2 u f(u)}+A_2''(u)=0\ , \nonumber\\
&& \frac{A_3(u) \left(\omega ^2-k^2 f(u)\right) R^3}{u^3 f(u)^2}+\frac{A_3'(u) \left(5 f(u)+2 u f'(u)\right)}{2 uf(u)}+A_3''(u)=0\ .
\end{eqnarray}
Clearly the $A_2$ and $A_3$ meson spectra are degenerate.
For convenience we again use the variable $x=u/u_T$. The solution close to the horizon is similar to the one obtained for the longitudinal mode (given in equation (\ref{eqt: Ehori})) and is given by
\begin{equation}\label{eqt: a2thori}
A_2(x)=C_T {\rm Exp}\left(-i\tilde{\omega}\log(x-1)\right)\ ,
\end{equation}
where $C_T$ is an yet undetermined constant. At small values of $\omega$ and $k$ (compared to the background temperature $T$) the solution for the transverse modes can be obtained to be
\begin{eqnarray}\label{eqt: a2t}
A_2(x)=N_2-\frac{1}{3}N_1\log\left[\frac{x^{3/2}+1}{x^{3/2}-1}\right]+\ldots \ ,
\end{eqnarray}
where $N_1$ and $N_2$ are constants of integration. 

Now expanding the solution in eqn. (\ref{eqt: a2t}) near the horizon and matching with the solution in eqn. (\ref{eqt: a2thori}) gives
\begin{eqnarray}
&&  N_2 -\frac{1}{3} N_1  \log \left(\frac{4}{3}\right)=C_T\ , \nonumber\\
&& \frac{1}{3}N_1 =-i C_T \tilde{\omega}\ .
\end{eqnarray}
Normalizability requires us to set $N_2=0$. However, we do not get a consistent solution for $N_1$ from the conditions above. From the equations of motion in (\ref{eqt: a23t}), it can be checked numerically (see also ref.~\cite{Evans:2008tv}) that the normalizability condition cannot be satisfied in the hydrodynamic limit (namely when $\tilde{\omega}\ll 1$ and $\tilde{k}\ll 1$). Therefore we can conclude that the transverse mode does not have any solution compatible with the limit $\tilde{\omega}\ll 1$ and $\tilde{k}\ll 1$ in accordance with the results obtained in, e.g. ref.~\cite{Kovtun:2005ev}.

\renewcommand{\theequation}{C-\arabic{equation}}
\setcounter{equation}{0}  
\section*{Appendix C}  

Here we outline the general variable changes to obtain the corresponding Schr\"{o}dinger equation from the original equation for fluctuation modes. Without any loss of generality we can write the equation for any fluctuation mode as follows
\begin{eqnarray}\label{eqt: seom}
a_1(u)f_\chi''(u)+a_2(u)f_\chi'(u)+\omega^2a_3(u)f_\chi(u)=0\ ,
\end{eqnarray}
where, $a_1(u)$, $a_2(u)$ and $a_3(u)$ are known functions and prime denotes derivative with respect to $u$, $f_{\chi}(u)$ denotes the radial profile of the fluctuation mode and $\omega$ is the corresponding oscillatory frequency (in the finite temperature case this is the quasinormal mode). Let us rewrite $f_\chi(u)=\sigma(u)g(u)$ with 
\begin{eqnarray}\label{eqt: schg}
\frac{\sigma'(u)}{\sigma(u)}=-\frac{1}{2}\left[\frac{a_2(u)}{a_1(u)}+\frac{1}{2}\left(\frac{a_1(u)}{a_3(u)}\right)\partial_u\left(\frac{a_3(u)}{a_1(u)}\right)\right]\ .
\end{eqnarray}
The Schr\"{o}dinger equation is then obtained to be
\begin{eqnarray}\label{eqt: scsch}
&& \sqrt{\frac{a_1(u)}{a_3(u)}}\partial_u\left(\sqrt{\frac{a_1(u)}{a_3(u)}}\left(\partial_u g(u)\right)\right)+\omega^2 g(u)-V_s(u) g(u)=0\ ,\quad {\rm where}\nonumber\\
&& V_s(u)=\frac{1}{B(u)^2}\left[\frac{1}{4}\left(A(u)+\frac{B'(u)}{B(u)}\right)^2+\frac{1}{2}\partial_u\left(A+\frac{B'(u)}{B(u)}\right)+\frac{A(u)}{2}\left(A(u)+\frac{B'(u)}{B(u)}\right)\right]\ , \nonumber\\
&& A(u)=\frac{a_2(u)}{a_1(u)}\ , \quad B(u)=\left(\frac{a_3(u)}{a_1(u)}\right)^{1/2}\ .
\end{eqnarray}
A more conventional form of the Schr\"{o}dinger equation can be obtained after changing variables to ``tortoise" coordinate $d\tilde{u}=B(u) du$; the horizon is then located at $\tilde{u}\to\infty$. The main goal of this exercise is to obtain the potential $V_S(u)$ and extract qualitative features of the meson spectrum. 

Here we have explicitly assumed that the fluctuation does not have any momentum mode. The fluctuation equation for the meson having a momentum can also be recast in the form of a Schr\"{o}dinger equation. The changes of variables are exactly similar as already mentioned in eqn. (\ref{eqt: scsch}), but the effective potential receives a positive contribution coming from the momentum 
\begin{eqnarray}
V_S(u,k)=V_S(u,0)+k^2\left(\frac{|a_4|}{a_3}\right)\ ,
\end{eqnarray}
where the left hand side of the equation (\ref{eqt: seom}) is now accompanied by another term of the form $-|a_4(u)|k^2 f_{\chi}$, the relative negative sign is due to the fact that $k$-corresponds to a spatial oscillation.

\renewcommand{\theequation}{D-\arabic{equation}}
\setcounter{equation}{0}  
\section*{Appendix D}  

Here we would like to solve a toy problem to study the effect of the magnetic field on meson dissociation. We implicitly assume the framework where we can have two equal charges at the two ends of the string and therefore analyze the symmetric configuration only. Rindler spacetime provides an useful arena where many qualitative (if not quantitative) features of an event horizon can be realized in a simple set up. Such study has been previously carried out in ref.~\cite{Peeters:2007ti} without any background field. The main result of this exercise is that the high spin meson dissociates once a critical value of the acceleration is reached. This acceleration is in turn determined by the angular momentum of the meson.

There are two ways to physically interpret the results of such pursuits (e.g., see ref.~\cite{Peeters:2007ti} and ref. \cite{Berenstein:2007tj}). We consider accelerating the rotating string in one of the space-like directions and therefore a Rindler horizon forms in the four dimensional spacetime. Alternatively, we can consider accelerating the string in the holographic direction and therefore a bulk spacetime horizon forms which sets the temperature of the dual gauge theory. To begin with we take the former point of view.

The metric for the Rindler space can be written as
\begin{eqnarray}\label{eqt: rmet}
ds^2=-\xi^2\kappa^2d\eta^2+d\xi^2+d\rho^2+\rho^2d\phi^2\ ,
\end{eqnarray}
where $\{\rho,\phi\}$-plane represents the plane where we would rotate the string. To add a background magnetic field we consider similar potential $A_\phi$ given before. Now the ansatz for the string in terms of its worldvolume coordinates $\{\tau,\sigma\}$ is given by
\begin{eqnarray}\label{eqt: ransatz}
\eta=\tau\ , \quad \xi=\xi(\sigma)\ ,\quad \rho=\rho(\sigma)\ , \quad \phi=\omega \tau\ .
\end{eqnarray}
This ansatz implies that the direction of acceleration of the string is orthogonal to the direction of its rotation. We can readily see that there are two conserved charges associated with the string, namely the angular momentum $J$, which is associated with the rotation along $\phi$-direction and the boost charge, which is associated with the translation in $\eta$-direction. 

The Nambu-Goto action for the string now will be accompanied by a boundary term exactly similar to the term $\Delta S_B$ given in eqn. (\ref{eqt: action}). Therefore the full action is given by
\begin{eqnarray}
S=\frac{1}{2\pi\alpha'}\int d\tau d\sigma \sqrt{(\xi^2\kappa^2-\omega^2\rho^2)(\xi'^2+\rho'^2)}+\Delta S_B\ .
\end{eqnarray}
Varying this action we can obtain the equation of motion for the string profile and the boundary conditions we should impose. In absence of any external field this exercise has been carried out in details in ref.~\cite{Peeters:2007ti}, which we do not repeat here. The presence of the magnetic field enforces us to impose the following boundary condition
\begin{eqnarray}\label{eqt: bcr}
\left. \pi_\rho\right|_{\pm \ell/2}=\left.\frac{\partial L}{\partial \rho'}\right|_{\pm \ell/2}=-B\rho\omega\ ,
\end{eqnarray}
where $\ell$ is the distance between the two end points of the string. The other boundary term corresponding to the variation of $\xi$ is satisfied imposing the Dirichlet boundary condition $\delta\xi=0$. This corresponds to a situation where the end points move with a constant acceleration $a=\xi(\pm\ell/2)^{-1}$.

Now choosing a gauge $\rho=\sigma$ we can solve\footnote{The analytical solution of the equations of motion obtained from the Nambu-Goto action remains the same as in ref.~\cite{Peeters:2007ti}. We refer to ref.~\cite{Peeters:2007ti} for further details about solving the equations of motion. We use the same analytical solution and impose the boundary condition given in (\ref{eqt: bcr}) to obtain the results in eqn. (\ref{eqt: rsol}).} the equation of motion to obtain the profile for the string given by
\begin{eqnarray}\label{eqt: rsol}
\xi(\rho)&=&\frac{C}{\kappa}\cosh\left[\frac{\kappa}{\omega}\arcsin\left(\frac{\omega\rho}{C}\right)\right]\ \quad {\rm with} \quad C^2=\frac{\omega^2\ell^2}{4}(1+B^2)\ .\nonumber\\
 J&=&\frac{1}{2\pi\alpha'}\frac{\ell^2}{4}\left(1+B^2\right)\arctan\left(\frac{1}{B}\right)\ , \quad T_s=\frac{1}{2\pi\alpha'}\ .
\end{eqnarray}
As the notation suggests the parameter $J$ represents the angular momentum of the meson. 

As we have seen before, the presence of the background magnetic field does not change the equations of motion for the Nambu-Goto string. Therefore the functional form of the string profile given in equation (\ref{eqt: rsol}) is entirely determined by the Rindler geometry\cite{Peeters:2007ti}. We notice that the presence of the magnetic field does induce an effective length $\ell_{\rm eff}=\ell^2(1+B^2)$ by changing the boundary condition of the string end-points.

A non-zero value of $B$ brings about two key changes in the evaluation of the angular momentum. The profile of the string modifies in order to satisfy the boundary condition in eqn. (\ref{eqt: bcr}) and therefore the effective length $\ell_{\rm eff}$ arises in the formula in (\ref{eqt: rsol}). There is also an additional explicit contribution $\Delta J_B$ of the form shown in eqn. (\ref{eqt: ej}), which finally gives a functional factor of $\tan^{-1}(1/B)$ in the expression for the angular momentum.

\begin{figure}[!ht]
\begin{center}
\includegraphics[angle=0,
width=0.65\textwidth]{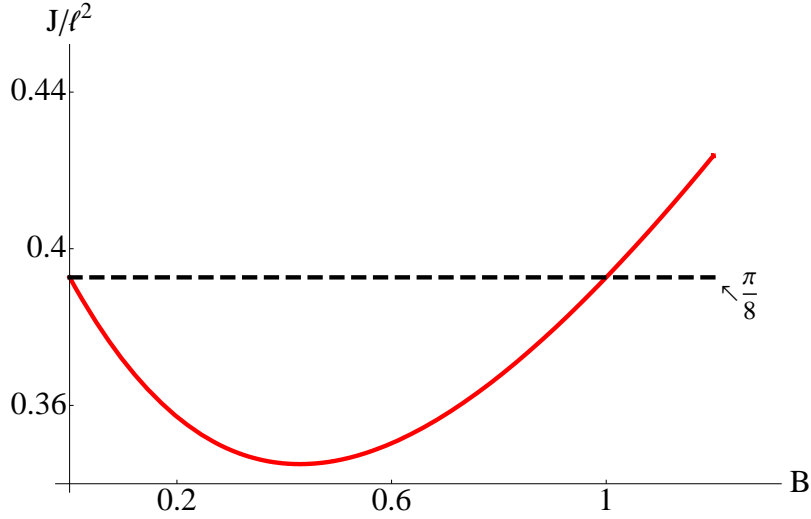}
\caption{\small The total angular momentum of the string in units of $(2\pi\alpha')$ in the background field. The horizontal dashed line represents the value of angular momentum of the string when there is no background magnetic field. We observe that the $B$-field initially reduces and then increases the angular momentum creating a local minima. However, we will observe that the stability of the meson is always enhanced by the $B$-field.}
\label{fig: jomegab}
\end{center}
\end{figure}

It can be shown that the explicit contribution of the of the magnetic field to the angular momentum, denoted by $\Delta J_B$, is actually cancelled by a term coming from integrating over the string profile. The net resulting angular momentum therefore can be lower than the angular momentum of the string in absence of the field. This is pictorially summarised in figure \ref{fig: jomegab}.

To find the maximum acceleration we identify $a^{-1}=\xi(\ell/2)$ and look for minima of the right hand side of the equation as a function of the parameter $C$. For now we content ourselves with only positive values of $B$, i.e. when the string angular momentum and the background field are parallel. It is a straightforward exercise to show that such minima corresponds to values of $C$ which satisfy the following equation (obtained from taking the first derivative of $\xi$ and setting it equal to zero at $\rho=\pm \ell/2$)
\begin{eqnarray}\label{eqt: solve}
&&x=m\tanh\left(\frac{m}{x}\right)\ , \quad {\rm where}\nonumber\\
&&x=\frac{C}{\kappa}\frac{1}{\sqrt{2\pi\alpha' J}}\ , \quad m=\arcsin\left(\frac{1}{\sqrt{1+B^2}}\right)\left[\arctan\left(\frac{1}{B}\right)\right]^{-1/2}\ .
\end{eqnarray}
Clearly the parameter $B$ generates a set of such values of $C$ which satisfy the above equation and therefore promotes the maximum value of acceleration to a function of the magnetic field. Differentiating the eqn. in the first line of (\ref{eqt: solve}) and using identities for the trigonometric hyperbolic functions we obtain that the family of roots of the equation can be simply given by the relation $x=\alpha m$ where $\alpha=0.834$ is a constant.

\begin{figure}[!ht]
\begin{center}
\includegraphics[angle=0,
width=0.65\textwidth]{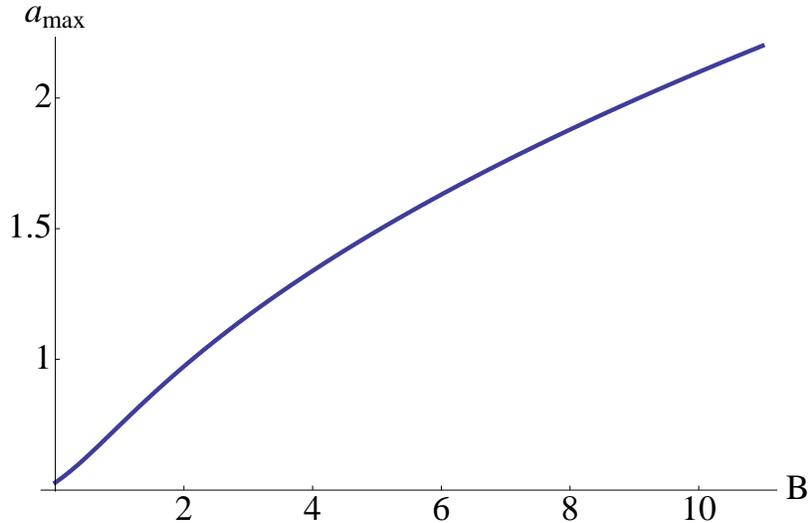}
\caption{\small The maximum acceleration in units of $\sqrt{2\pi\alpha' J}$ as a function of the external magnetic field.}
\label{fig: amaxb}
\end{center}
\end{figure}

Using these roots the maximum acceleration can be obtained as a function of the background magnetic field which is given by
\begin{eqnarray}
a_{\rm max}\propto \sqrt{\arctan\left(\frac{1}{B}\right)}\frac{1}{\arcsin\left(\frac{1}{\sqrt{1+B^2}}\right)}\ .
\end{eqnarray}
The functional dependence is pictorially shown in figure \ref{fig: amaxb}.
We observe that the maximum acceleration grows almost linearly with the background field.

Alternatively we could imagine the Rindler space to be the near-horizon approximation of a blackhole background and the acceleration is along the holographic coordinate (we make the identification that $\xi=u$, where $u$ is the radial coordinate in the holographic set up). The parameter $\kappa$ (in eqn. (\ref{eqt: rmet})) in this case represents the surface gravity and sets the temperature of the dual gauge theory to be $T=\kappa/(2\pi)$. 

Now we imagine the string end points to sit on the flavour brane at some value of $\xi_0=U_0$, which fixes the constituent quark mass; and we consider a similar ansatz for the string profile as presented in eqn. (\ref{eqt: ransatz}). The interpretation of the set up is different from the earlier one; however this brings about no change as far as the mathematics is concerned. We therefore again obtain the string profile as given by eqn. (\ref{eqt: rsol}).

However in this case the physical parameter that captures the dissociation of the meson is the size $\ell$ (which sets the angular momentum $J$ via the relation in eqn. (\ref{eqt: rsol})) of it. The length is obtained to be
\begin{eqnarray}\label{eqt: rlen}
\ell=\frac{2C}{\kappa}\frac{{\rm arccosh}\left(\kappa U_0/C\right)}{\sqrt{1+B^2}}\left[\arcsin\left(1/\sqrt{1+B^2}\right)\right]^{-1}\ .
\end{eqnarray}
We would hope to see that for a given value of $U_0$, there exists a maximum admissible size of the spinning meson beyond which it dissociates. 

We can treat as before the parameter $C$ to be independent with respect to which we will consider maximizing the length function $\ell$. From eqn. (\ref{eqt: rlen}) it is clear that the external magnetic field does not play any role to determine the value of $C$ corresponding to the maximum of $\ell$; however it determines the the function $\ell_{\rm max}(B)$ and therefore also $J_{\rm max}(B)$ to be given by
\begin{eqnarray}\label{eqt: ljmax}
\ell_{\rm max}\propto \frac{1}{\sqrt{1+B^2}}\left[\arcsin\left(1/\sqrt{1+B^2}\right)\right]^{-1}\ , \quad J_{\rm max}\propto \arctan\left(\frac{1}{B}\right) \left[\arcsin\left(1/\sqrt{1+B^2}\right)\right]^{-2}\ .\nonumber\\
\end{eqnarray}
where the constants of proportionality depends on the background temperature $T$ and the parameter $U_0$.

\begin{figure}[!ht]
\begin{center}
\subfigure[] {\includegraphics[angle=0,
width=0.45\textwidth]{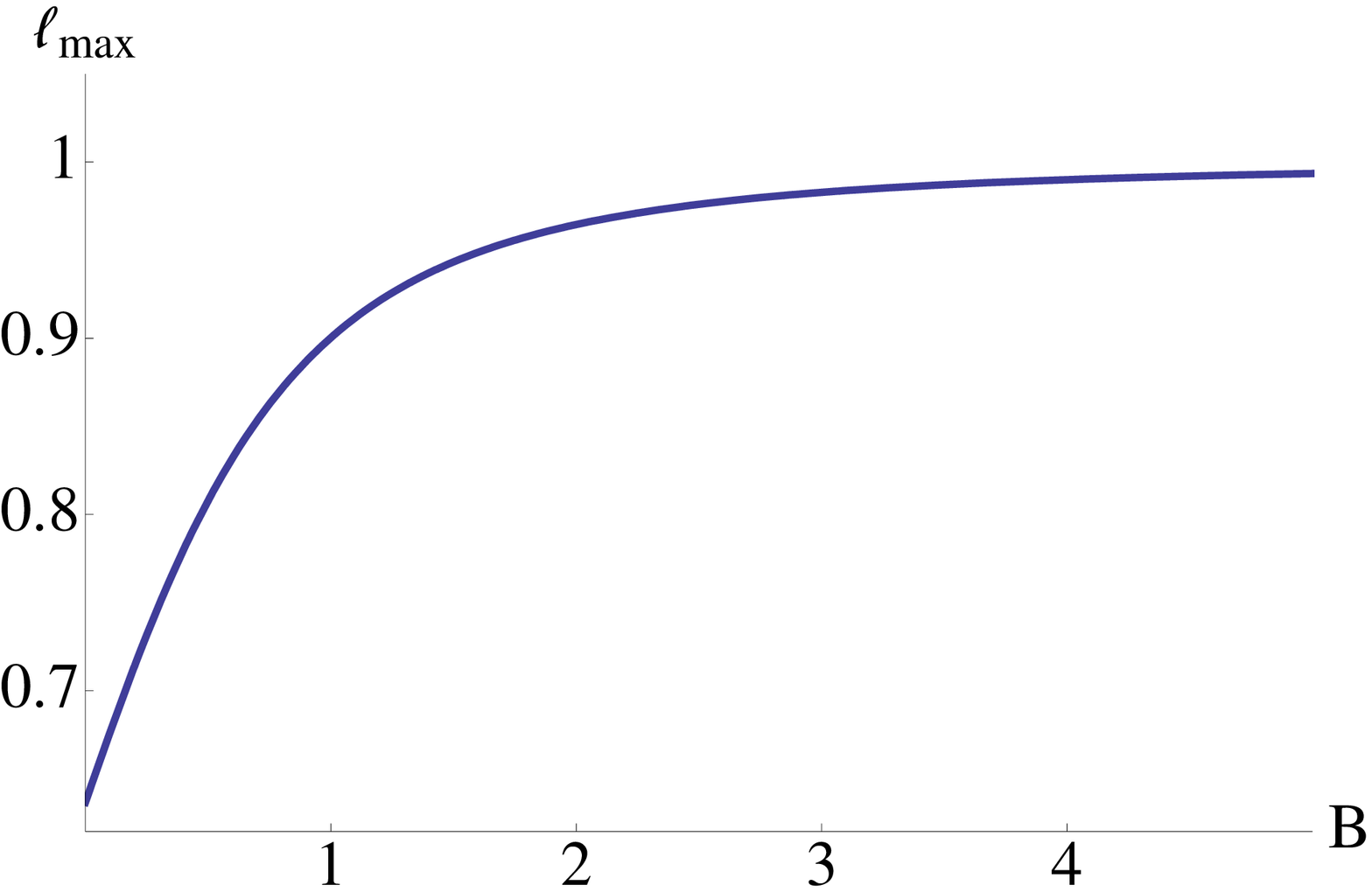} \label{fig: rlmax}}
\subfigure[] {\includegraphics[angle=0,
width=0.45\textwidth]{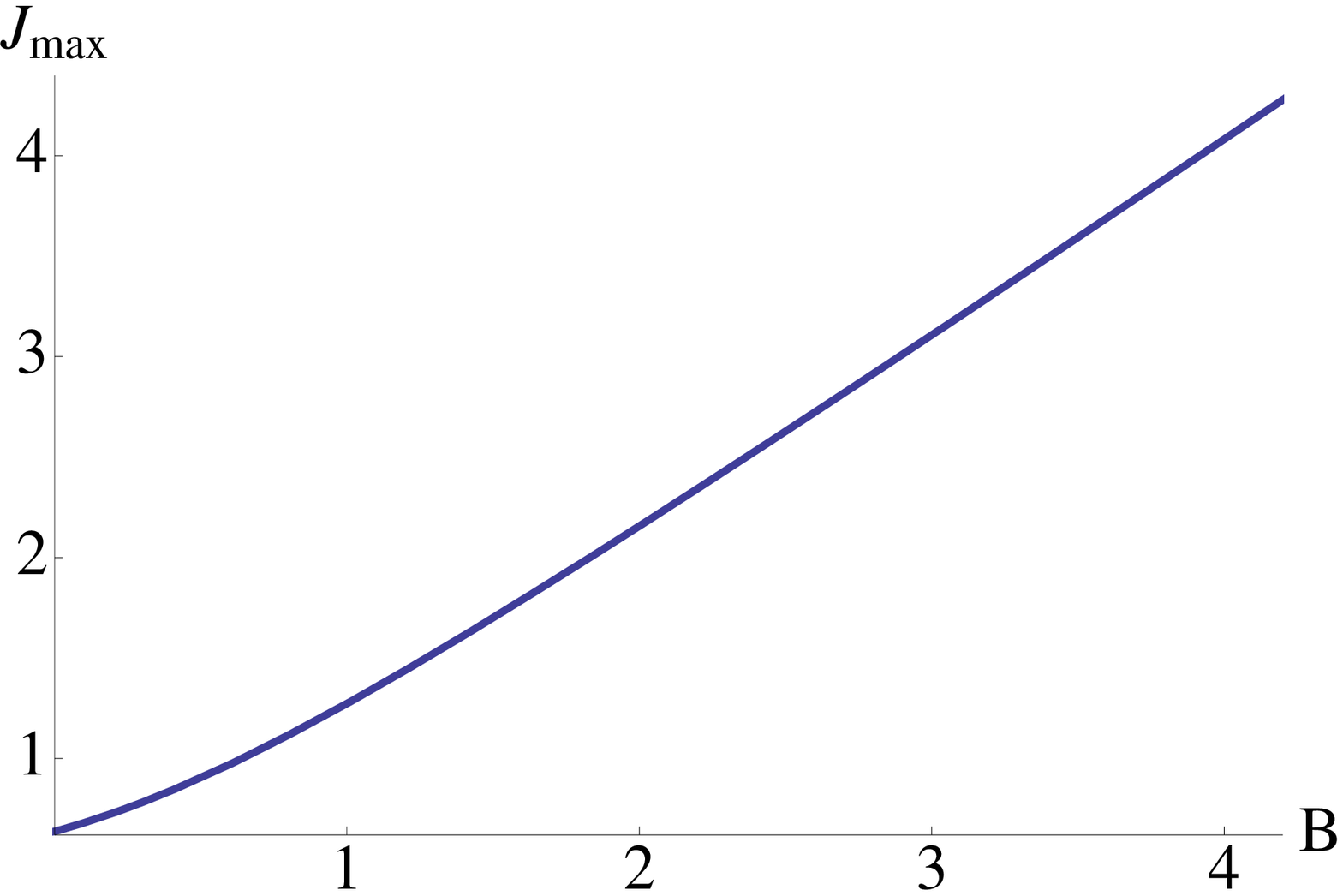} \label{fig: rjmax}}
\caption{\small We plot the two functions given in eqn. (\ref{eqt: ljmax}). The constants of proportionality have not been taken into account here.}
\end{center}
\end{figure}

For illustrative purpose we have plotted the functional behaviour of $\ell_{\rm max}$ and $J_{\rm max}$ in figure \ref{fig: rlmax} and \ref{fig: rjmax} respectively. We see that the maximum length of the meson saturates an upper bound at high enough values of the magnetic field; however no such saturation is present in the maximum angular momentum of the meson.

For the background considered in the main text the Rindler spacetime does emerge when we zoom in the region close to the horizon\cite{Frolov:2006tc, Mateos:2006nu}. The toy model here can therefore be identified with a corresponding study of high spin mesons in the overheated phase. So if we want to make a direct connection between the Rindler background dynamics to the specific system studied in the text, we should remember that such a background appears precisely in the over-heated phase of the system, where any small fluctuation is likely to destroy the meson and drive the over-heated phase to the melted phase.

\providecommand{\href}[2]{#2}\begingroup\raggedright

\end{document}